\documentclass[%
 reprint,
 amsmath,amssymb,
 aps,
]{revtex4-2}

\usepackage{graphicx}% Include figure files
\usepackage{dcolumn}% Align table columns on decimal point
\usepackage{bm}% bold math
\usepackage{hyperref}
\usepackage{amsmath}
\usepackage{hyperref}
\hypersetup{
    colorlinks=true,        % false: boxed links; true: colored links
    linkcolor=blue,         % color of internal links (change box color with linkbordercolor)
    citecolor=blue,         % color of links to bibliography
    filecolor=blue,         % color of file links
    urlcolor=blue           % color of external links
}

\begin{document}

\preprint{APS/123-QED}

\title{Determining the Dielectric Constant of Solid/Liquid Interfaces}% Force line breaks with \\

\author{Somaiyeh Dadashi$^1$, Narendra M. Adhikari$^2$, Hao Li$^1$, Stefan M. Piontek$^1$, Zheming Wang$^2$, Kevin M. Rosso$^2$, Eric Borguet$^1$}
\email[Correspondence email address: ]{eborguet@temple.edu}% Your name
\affiliation{$^1$Department of Chemistry, Temple University, Philadelphia, PA 19122, USA\\$^2$Pacific Northwest National Laboratory, Richland, WA 99354, USA}

\begin{abstract}
The dielectric constant ($\varepsilon^{\prime}$) of interfacial water is an important parameter, but its measurement has posed challenges, and no consensus    has been reached on a generalized expression.   We derived a formula for $\varepsilon^{\prime}$ of a buried interface using the slab model for a half-solvated sphere: $ \varepsilon^{\prime}=\varepsilon_1 \varepsilon_2\left(\varepsilon_2-\varepsilon_1+6\right) / 2\left(2 \varepsilon_2+\varepsilon_1\right)
$, where $\varepsilon_1$ and $\varepsilon_2$  are the dielectric constants of the solid and liquid phases, respectively.  We experimentally validated this expression using vibrational sum frequency generation and Fresnel factor calculations for interfaces of alumina with water ($ \mathrm{H_2O} $ and $ \mathrm{D_2O} $) and acetonitrile.  This fills an important knowledge gap in the description of the dielectric constant of interfaces.  
\end{abstract}

\maketitle

%\tableofcontents
\textit{Introduction.---} The unique properties of liquids at interfaces or under confinement arise from asymmetric forces and molecular interactions that differ from those in bulk \cite{sit2013, eisenthal1996}.  Notably, the reduced rotational freedom of water dipoles near surfaces can lead to a decrease in the dielectric constant  at the interface \cite{fumagalli2018, zhu2020}, influencing surface interactions, mass transport, ionic adsorption \cite{wang2019sodium}, and chemical equilibria \cite{boyes2020}, which in turn impact atmospheric chemistry \cite{jimenez2009, yu2021, zhong2018} as well as various chemical and biological processes \cite{deissenbeck2023,bagchi2013}.  For example, the interfacial electric fields, which according to Coulomb’s law inversely depend on the interfacial dielectric constant, are believed to be the driving force of the high activity of interfacial water environments in “on-droplet” chemistry \cite{lee2019, xiong2020}.  Therefore, understanding the dielectric constant of the interface is crucial due to its significant contributions to many interfacial phenomena \cite{sit2013, teschke2001, chiang2022, loche2020,munoz-santiburcio2017}.

Multiple experimental and theoretical approaches \cite{xiao2019}, including plasmonic sensors \cite{zhan2020}, atomic force microscopy \cite{fumagalli2018,teschke2001,teschke2005}, and classical molecular dynamics (MD) simulations of water in charged nanopores \cite{underwood2022}, have been employed to investigate the static and optical dielectric constants of water in interfacial regions or confined environments.   For example, the value of the static dielectric constant at the air/water interface has been reported to be $\sim$2 \cite{teschke2005}.    Additionally, local capacitance measurements have shown that the 2-3 layers of water confined between thin channels of hexagonal boron nitride crystals have an out-of-plane dielectric constant of $\sim$2  \cite{fumagalli2018}.  There is qualitative agreement that these values are significantly smaller compared to the bulk water static dielectric constant of $\sim$80, and above its value of $\sim$1.8 at optical frequencies \cite{fumagalli2018,teschke2001,teschke2005}.

The dielectric constant, a frequency dependent quantity, has four distinct components: ionic transport, dipole reorientation, vibrational and electronic resonances \cite{othman2011}.  Under confinement or at interfaces, the reorientation of molecules is likely reduced so that vibrational and electronic responses should dominate the dielectric function \cite{fumagalli2018}.  Surface specific techniques, e.g., vibrational sum frequency generation (vSFG) spectroscopy, whose response depends on the dielectric constant of the medium, should be sensitive to the interfacial dielectric constant \cite{teschke2001,chiang2022}.  For example, the vSFG response is inherently sensitive to the interfacial dielectric environment, as the local dielectric function governs the strength of the electromagnetic fields (via Fresnel factors) at the interface \cite{shen1989}.  Consequently, the vSFG spectra can yield insight into the interfacial dielectric constant.

The dielectric constant at the air/water interface has been modeled considering the inhomogeneous interfacial dielectric profile using the Lorentz and the slab models \cite{chiang2022,wang2021}.  The Lorentz model assumes that the chromophores are completely solvated, resulting in $\varepsilon^{\star}=\varepsilon\left(\varepsilon^{\star}\right.$: interfacial dielectric constant and $\varepsilon$: dielectric constant of water) \cite{morita2018}.  The slab model, on the contrary, assumes that the interfacial molecules are partially solvated and treated as idealized hemispheres, leading to $\varepsilon^{\star}=\varepsilon(\varepsilon+5)(4\varepsilon+2)$ for the air/liquid interface \cite{zhuang1999}.  Both models are commonly used for liquid/vapor interfaces, where the dielectric constant of air is 1, and $\varepsilon^{\prime}$ is largely determined by the liquid medium's dielectric constant.  However, whether these formulas for the dielectric constant accurately describe the environment at buried, e.g., solid/liquid, interfaces is not clear.

We derived expressions for $\varepsilon^{\prime}$ for the solid/liquid interface based on the slab and Lorentz models.  We then devised an approach to test their validity through angle-dependent vSFG spectroscopy and Fresnel factor calculations using the $\alpha$-$ \mathrm{Al_2O_3} $(0001) surface in contact with $ \mathrm{H_2O} $, $ \mathrm{D_2O} $, and acetonitrile.  The results showed that the slab model more accurately estimates the interfacial dielectric constant.  We further compared the angle-dependent Fresnel factors using distinct $\varepsilon^{\prime}$ models, which enabled us to determine the interfacial dielectric constant, $\varepsilon^{\prime}$.  Our calculations, confirmed by our experiments, suggest that $\varepsilon^{\prime}$ can be defined using the expression that we derive for buried interfaces, $ \varepsilon^{\prime}=\varepsilon_1 \varepsilon_2\left(\varepsilon_2-\varepsilon_1+6\right) / 2\left(2 \varepsilon_2+\varepsilon_1\right)$  where $\varepsilon_1$ and $\varepsilon_2$ are the dielectric constants of solid and liquid phase, respectively, which generalizes the approach of Shen \emph{et al.}   \cite{zhuang1999}.  Although the majority of models discussed in the literature were not effective in estimating the interfacial dielectric constant, one formula, derived from classical MD simulation \cite{wang2017}, shows good agreement with our expression for $\varepsilon^{\prime}$, which further supports its validity.

\textit{Experimental methods.---} Details of experiments, including sample preparation, optical setup, spectrum normalization and other information, can be found in the Supplemental Material section I and II.  

\textit{Results and discussion.---}
Expressions for the dielectric constant of the interfacial region, $\varepsilon^{\prime}$, have been determined by several different approaches \cite{wang2017,shiratori2011}.  Morita \emph{et al.} used classical MD simulations to determine an expression, 
$\left(0.8 \sqrt{\varepsilon_2}+0.2 \sqrt{\varepsilon_1}\right)^2$, for the interfacial dielectric constant of the vapor ($\varepsilon_1$)/liquid ($\varepsilon_2$) interface \cite{wang2017}.  Shen \emph{et al.} derived an expression, $\varepsilon^*=\varepsilon_2\left(\varepsilon_2+5\right) /\left(4 \varepsilon_2+2\right)$, using the slab model to calculate local-field corrections for the monolayer/air interface \cite{zhuang1999}.  The applicability of $\varepsilon^*$ expression to buried, e.g., solid/liquid, interfaces remains unexplored and debatable.  
For example, at solid/liquid interfaces, where the solid phase, such as $\mathrm{Al_2O_3} $, has a higher dielectric constant ($\varepsilon_1$), with $\varepsilon$ $\approx$ 2.9 at 3350 $ \mathrm{cm^{-1}} $, compared to the liquid phase ($\varepsilon_2$), e.g., water with $\varepsilon$ = 1.74 $+$ i0.75, the impact of both $\varepsilon_1$ and $\varepsilon_2$ must be considered in the calculation of the interfacial dielectric constant \cite{malitson1962}.

Following classical electrodynamics, we used two approaches, the Lorentz and slab models, to calculate the interfacial dielectric constant \cite{bottcher1973}.  The slab model, which considers a half-solvated shell for molecules at the interface, shows good agreement with our results.  The Lorentz model, however, did not show good agreement with the experiments as it overestimated $\varepsilon^{\prime}$ when compared to the experimental results (Further information comparing the two approaches can be found in the Supplemental Material section III and IV).  We believe the success of the half-solvated model is due to its more accurate representation of interfacial molecules, which, compared to bulk molecules, should lack part of their solvation shell \cite{eftekhari-bafrooei2010,lesnicki2020}.  We used an approach similar to Shen \emph{et al.} \cite{zhuang1999}, who derived an equation for the solid/air interface dielectric constant, to calculate $\varepsilon^{\prime}$ at the solid/liquid interface considering that the dielectric constant of the solid medium, ($\varepsilon_1$), is more than 1.  An external field induces charges on a dielectric sphere ($\varepsilon_1$) 
\begin{figure}[h]
\includegraphics{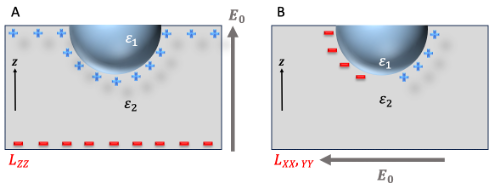}% Here is how to import EPS art
\caption{\label{FIG1} Schematic representation of the slab model for a hemisphere interacting with an external electric field $E_0$ along A) the Z-axis and B) the X or Y-axes.}
\end{figure} in a dielectric medium of different dielectric constant ($\varepsilon_2$) (FIG. \ref{FIG1}A), generating an electric field due to the polarization of the medium.  As a result, the local electric field is amplified through additional dipole fields generated by the polarization on the sphere in the X, Y and Z directions (FIG. \ref{FIG1}A and \ref{FIG1}B) \cite{shiratori2011}.

We calculated the induced local fields perpendicular, $E_\bot$ FIG.( \ref{FIG1}A), and parallel, $E_\parallel$ (FIG \ref{FIG1}B), to the surface using the approach for a dielectric hemisphere of dielectric constant ($\varepsilon_1$) in a medium with a different dielectric constant ($\varepsilon_2$) \cite{bottcher1973}.  Based on Shen \emph{et al.}, the local field at the interface is the sum of the external field, $E_0$, and the field generated through polarization inside the hemisphere, which in our case represents a half solvated molecule, and the surrounding semi-infinite medium \cite{zhuang1999}.   However, due to the isotropic symmetry of the medium, the dipole field originating from the polarization in the bulk is zero \cite{zhuang1999}.  $E_\bot$ and $E_\parallel$, can be described using Eq.~\ref{eq:one} and Eq.~\ref{eq:two} (See Supplemental Material section III for further details): 

\begin{eqnarray}
E_{\perp}=E_0\left(\frac{\varepsilon_2-\varepsilon_1+3}{3}\right)
\label{eq:one}
\end{eqnarray}
%(\ref{eq:one})

\begin{eqnarray}
E_1=E_0\left(\frac{2 \varepsilon_2+\varepsilon_1}{3 \varepsilon_2}\right)
\label{eq:two}
\end{eqnarray}
%(\ref{eq:two})
The local field applied to the molecules at the interface, which in our case is the half-solvated hemisphere, is proportional to the external electric field, $E_0$.  The local fields can be defined through the local electric field correction factors, discussed below in more detail, by knowing the dielectric functions of the media and the interfacial region as well as the angles of incidence of the input laser beams.  These correction factors are embedded in the vSFG response, enabling this technique to be used for determining the unknown interfacial dielectric function \cite{piontek2023}.

Sum frequency generation is a surface-specific second-order nonlinear technique \cite{wang2015,piontek2023}. Within the electric dipole approximation, all second-order processes are forbidden in isotropic media, as the effective second-order susceptibility, $\chi_{\text {eff }}^{(2)}$, vanishes due to symmetry constraints \cite{wang2015}.  However, the intrinsically non-centrosymmetric interface results in a  non-vanishing $\chi_{\text {eff }}^{(2)}$  that enables the generation of a vSFG response (Eq.~\ref{eq:three}) \cite{wang2015}. 
\begin{equation}
\begin{aligned}
I\left(\omega_{\text{v SFG}}\right)
&= \left|\chi_{NR}^{(2)} e^{i \varphi_{NR}} + \sum_v \frac{A_\nu}{\omega_{\text{IR}} - \omega_\nu + i \Gamma_\nu}\right|^2 I_{\text{Vis}} I_{\text{IR}}
\end{aligned}
\label{eq:three}
\end{equation}
\\
where \(A_{\nu}\) is the amplitude, \(\Gamma_{\nu}\) is the damping coefficient, \(\omega_{\nu}\) represents the central frequency of the (\(\nu_{th}\) vibrational mode, and \(\varphi_{NR}\) is the phase of the second order non-resonant (\(\chi_{NR}^{(2)}\)) nonlinear susceptibilities \cite{shiratori2011}.  The vSFG response can be measured under SSP (S- for vSFG, S- for visible, and P- for IR), SPS, PSS and PPP polarization combinations, each probing different components of $\chi_{\text{eff}}^{(2)}$ tensor that can be expressed by macroscopic elements $\chi_{IJK}^{(2)}$ (I, J, K = X, Y, Z) through the following equations, Eq.~\ref{subeq:1-1}- Eq.~\ref{subeq:4}, with S referring to s-polarized light and P to p-polarized light, where the electric field is oriented perpendicular and parallel to the plane of incidence, respectively \cite{shiratori2011, busson2023sum}. 

\begin{subequations}
\begin{equation}
\begin{aligned}
\chi_{\text{eff,\;SSP}}^{(2)} &= L_{YY}\left(\omega_{SFG}\right) L_{YY}\left(\omega_{VIS}\right) L_{ZZ}\left(\omega_{IR}\right) \\
&\quad \sin \theta_{IR}\; \chi_{YYZ}
\end{aligned}
\label{subeq:1-1}
\end{equation}
\vspace{-1.5em}
\begin{equation}
\begin{aligned}
\chi_{\text{eff,\;SPS}}^{(2)} &= L_{YY}\left(\omega_{SFG}\right) L_{ZZ}\left(\omega_{VIS}\right) L_{YY}\left(\omega_{IR}\right) \\
&\quad \sin \theta_{VIS}\; \chi_{YZY}
\end{aligned}
\label{subeq:2}
\end{equation}
\vspace{-1.5em}
\begin{equation}
\begin{aligned}
\chi_{\text{eff,\;PSS}}^{(2)} &= L_{ZZ}\left(\omega_{SFG}\right) L_{YY}\left(\omega_{VIS}\right) L_{YY}\left(\omega_{IR}\right) \\
&\quad \sin \theta_{SFG}\; \chi_{ZYY}
\end{aligned}
\label{subeq:3}
\end{equation}
\vspace{-1.5em}
\begin{equation}
\begin{aligned}
\chi_{\text{eff,\;PPP}}^{(2)} &= -L_{XX}\left(\omega_{SFG}\right) L_{XX}\left(\omega_{VIS}\right) L_{ZZ}\left(\omega_{IR}\right) \\
&\quad \cos \theta_{SFG}\; \cos \theta_{VIS} \; \sin \theta_{IR}\; \chi_{XXZ} \\
&\quad -L_{XX}\left(\omega_{SFG}\right) L_{ZZ}\left(\omega_{VIS}\right) L_{XX}\left(\omega_{IR}\right) \\
&\quad \cos \theta_{SFG} \; \sin \theta_{VIS} \; \cos \theta_{IR}\; \chi_{XZX} \\
&\quad +L_{ZZ}\left(\omega_{SFG}\right) L_{XX}\left(\omega_{VIS}\right) L_{XX}\left(\omega_{IR}\right) \\
&\quad \sin \theta_{SFG} \; \cos \theta_{VIS} \; \cos \theta_{IR}\; \chi_{ZXX} \\
&\quad +L_{ZZ}\left(\omega_{SFG}\right) L_{ZZ}\left(\omega_{VIS}\right) L_{ZZ}\left(\omega_{IR}\right) \\
&\quad \sin \theta_{SFG} \; \sin \theta_{VIS}\;  \sin \theta_{IR}\; \chi_{ZZZ}
\end{aligned}
\label{subeq:4}
\end{equation}
\end{subequations}

where $\theta$ is the incident angle with respect to the surface normal, and the $L_{I I}(I=X, Y, Z)$ are the Fresnel factors that describe the local field corrections \cite{wang2015}.

When an external electric field is applied to the interface, the medium becomes polarized, consequently altering the electric field at the interface \cite{shiratori2011}. The Fresnel factors, elements of a 3 $\times 3$ tensor, are local field correction factors $\left(L_{I I}: L_{X X}, L_{Y Y}, L_{Z Z}\right)$ that account for changes in the electric fields at interfaces \cite{shiratori2011}. The Fresnel factors can be calculated (Eq.~\ref{subeq:5a} - Eq.~\ref{subeq:5c}) knowing the dielectric constants of medium $1\left(\varepsilon_1\right)$, medium $2\left(\varepsilon_2\right)$ and the interface $\left(\varepsilon^{\prime}\right)$ as well as the angles of the incident beams \cite{shiratori2011}: 

\begin{subequations}
\begin{equation}
\begin{aligned}
L_{XX}\left(\omega_I\right) &= \frac{2 \varepsilon_1\left(\omega_I\right) \cos \gamma_I}{\varepsilon_1\left(\omega_I\right) \cos \gamma_I + \varepsilon_2\left(\omega_I\right) \cos \theta_I}
\end{aligned}
\label{subeq:5a}
\end{equation}
\vspace{-1.5em}
\begin{equation}
\begin{aligned}
L_{YY}\left(\omega_I\right) &= \frac{2 \varepsilon_1\left(\omega_I\right) \cos \theta_I}{\varepsilon_1\left(\omega_I\right) \cos \theta_I + \varepsilon_2\left(\omega_I\right) \cos \gamma_I}
\end{aligned}
\label{subeq:5b}
\end{equation}
\vspace{-1.5em}
\begin{equation}
\begin{aligned}
L_{ZZ}\left(\omega_I\right) &= \frac{2 \varepsilon_2\left(\omega_I\right) \cos \theta_I}{\varepsilon_1\left(\omega_I\right) \cos \gamma_I + \varepsilon_2\left(\omega_I\right) \cos \theta_I} \frac{\varepsilon_1\left(\omega_I\right)}{\varepsilon^{\prime}\left(\omega_I\right)}
\end{aligned}
\label{subeq:5c}
\end{equation}
\end{subequations}
\\
where the $\theta_I$ represent the angles of incidence for the input beams, and the $\gamma_{\mathrm{I}}$ are the angles of the transmitted waves calculated using Snell's law \cite{lobau1997}. The interfacial dielectric constant is buried in the Fresnel factor, $L_{Z Z}\left(\omega_l\right)$, offering the opportunity to determine the interfacial dielectric constant through angle-dependent vSFG experiments.

The electric field components at the interface, $E_{\perp}$ and $E_{\|}$, can be determined using the Fresnel equations \cite{shiratori2011}. Knowing that $L_{Z Z} / L_{Y Y}=\varepsilon_1 / \varepsilon^{\prime}, L_{X X}=L_{Y Y}=E_{\perp} / E_0$ and $\mathrm{L}_{Z Z}=$ $E_{\|} / \mathrm{E}_0$ we have \cite{zhuang1999} (See Supplemental Material section III and Eq. S13 for further details):

\begin{equation}
\begin{aligned}
& \varepsilon^{\prime}=\frac{\varepsilon_1 \varepsilon_2\left(\varepsilon_2-\varepsilon_1+6\right)}{2\left(2 \varepsilon_2+\varepsilon_1\right)}
\end{aligned}
\label{eq:seven}
\end{equation}

At optical frequencies, we typically use the refractive index as per Maxwell's relation, $n=\sqrt{\varepsilon}$, and therefore \cite{teschke2001,chiang2022}:
\begin{equation}
\begin{aligned}
n^{\prime}=\sqrt{\frac{n_1^2 n_2^2\left(n_2^2-n_1^2+6\right)}{2\left(2 n_2^2+n_1^2\right)}}
\end{aligned}
\label{eq:eight}
\end{equation}

To validate the expression that we derived for $\varepsilon^{\prime}$, we collected angle-dependent vSFG spectra under three different experimental geometries that were selected, considering the critical angles at the solid/liquid interfaces. The vSFG experiments were performed at two total internal reflection (TIR) geometries (i.e., with angle-of-incidence higher than the critical angle for the IR and visible beams at the solid/liquid interface; Table S1 in the Supplemental Material section V), TIR1: $\theta_{\text {Vis }} \approx$ $63^{\circ}, \theta_{I R} \approx 57^{\circ}$, TIR2: $\theta_{V i s} \approx 54^{\circ}, \theta_{I R} \approx 60^{\circ}$ and in an "external" geometry, where the angles of incidence of the incoming beams are less than the critical angle: $\theta_{V i s} \approx 31^{\circ}, \theta_{I R} \approx 29^{\circ}$ (Further details can be found in the Supplemental Material section V). TIR geometries were selected for to improve the signal-to-noise ratio.

\begin{figure*}
\includegraphics{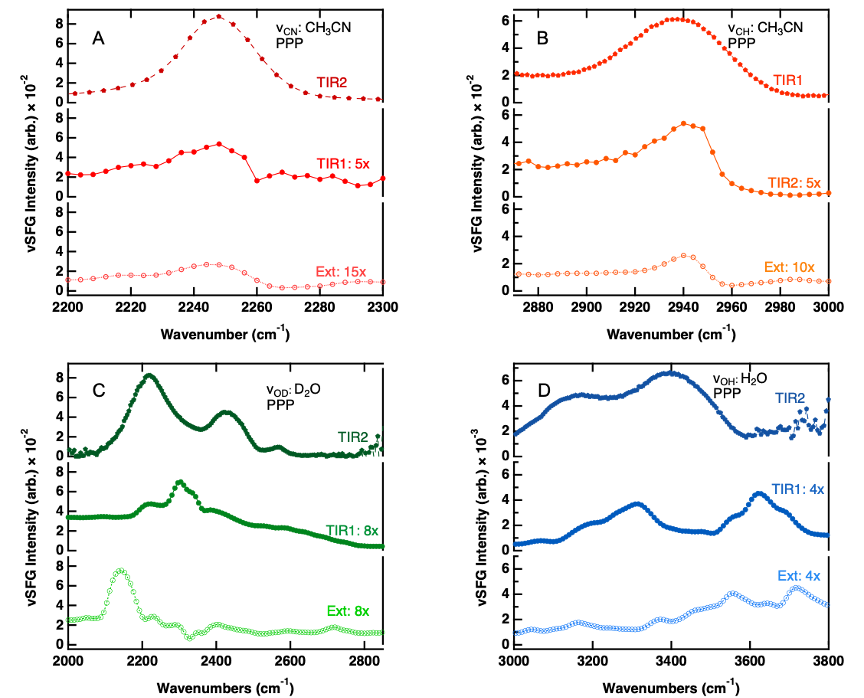}% Here is how to import EPS art
\caption{\label{fig2:wide} vSFG spectra for A) $\mathrm{C} \equiv \mathrm{N}$ stretch of $\left.\mathrm{CH}_3 \mathrm{CN}, \mathrm{B}\right) \mathrm{CH}$ stretch of $\left.\mathrm{CH}_3 \mathrm{CN}, \mathrm{C}\right) \mathrm{OD}$ stretch of $\mathrm{D}_2 \mathrm{O}$ and the $\mathrm{OH}$ stretch of $\mathrm{H}_2 \mathrm{O}$ at $\alpha{-}\mathrm{Al}_2\mathrm{O}_3(0001)$ interfaces. All of the experiments were performed using the PPP polarization combination in different geometries, TIR2 (Top), TIR1 (Middle), and Ext. (Bottom) in each panel.}
\end{figure*}

We performed vSFG experiments on aqueous/aluminum oxide/ interfaces, due to their abundance and importance in nature \cite{banuelos2023, piontek2023}.  Specifically, we investigated the $\mathrm{H}_2\mathrm{O}/\alpha{-}\mathrm{Al}_2\mathrm{O}_3(0001)$, and $\mathrm{D}_2\mathrm{O}/\alpha{-}\mathrm{Al}_2\mathrm{O}_3(0001)$ interfaces in the $\mathrm{OH}$ and the $\mathrm{OD}$ stretch regions, respectively, as well as the $\mathrm{CH}_3\mathrm{CN}/\alpha{-}\mathrm{Al}_2\mathrm{O}_3(0001)$ in the $\mathrm{C}=\mathrm{N}$ stretch and the $\mathrm{C}-\mathrm{H}$ stretch regions (FIG. \ref{fig2:wide}) interfaces. The vSFG spectra of acetonitrile in the $\mathrm{C} \equiv \mathrm{N}$ stretch and $\mathrm{CH}$ stretch region are similar to previous studies of this solid/liquid interface in that a single feature is observed near $\sim 2250-2260 \mathrm{~cm}^{-1}$ (FIG. \ref{fig2:wide}A) and near 2940-2950 cm ${ }^{-1}$ (FIG. \ref{fig2:wide}B) \cite{lock2002,rehl2018,rivera2013,wang2014,zhang1993,dadashi2024}.   We did not observe the Fermi resonance peak which arises from the coupling between the $\mathrm{C} \equiv \mathrm{N}$ stretch and the combination band ($\sim 2300 \mathrm{~cm}^{-1}$) of the $\mathrm{C}-\mathrm{C}$ stretching and $\mathrm{C}-\mathrm{H}$ bending modes since this mode is expected to lie at the edge of the IR profile used in our experiments \cite{dereka2021}. The presence of single narrow peaks in the $\mathrm{C} \equiv \mathrm{N}$ and the $\mathrm{CH}$ stretching spectral regions makes these features effective for benchmarking the dielectric constant at the interface \cite{rehl2018,bertie1997}.
\\
\indent As the experimental geometry changes from TIR2 to TIR1 and then to the external geometry, $I\left(\omega_{\nu S F G}\right)$ decreases for the $\mathrm{C} \equiv \mathrm{N}$ stretch and $\mathrm{CH}$ stretch of $\mathrm{CH}_3 \mathrm{CN}$ at liquid/$\alpha$-$\mathrm{Al}_2 \mathrm{O}_3(0001)$ interface.  The OD stretch $\mathrm{D}_2 \mathrm{O}$ (FIG. \ref{fig2:wide}C) and the $\mathrm{OH}$ of $\mathrm{H}_2 \mathrm{O}$ (FIG. \ref{fig2:wide}D) for TIR2 and TIR1 shows distinct peaks for strongly hydrogen bonded (OD $\sim 2200-2400 \mathrm{~cm}^{-1}$ and $\mathrm{OH} \sim 3000-3400 \mathrm{~cm}^{-1}$ ) and weakly hydrogen bonded ( $\mathrm{OD} \sim 2400-2700 \mathrm{~cm}^{-1}$ and $\mathrm{OH} \sim 3400-3700 \mathrm{~cm}^{-1}$ ) regions. For the external geometry, we observe an additional, relatively weak peak near $\sim 2730 \mathrm{~cm}^{-1}$ for $\mathrm{OD}$ stretch (FIG. \ref{fig2:wide}C) and near $\sim 3750 \mathrm{~cm}^{-1}$ for $\mathrm{OH}$ stretch (FIG. \ref{fig2:wide}D). These $\mathrm{OD}$ and $\mathrm{OH}$ stretch spectral characteristics suggest that distinct hydrogen bonding environments are probed in the external geometry compared to what is detected in the TIR geometries.

\begin{figure*}
\includegraphics{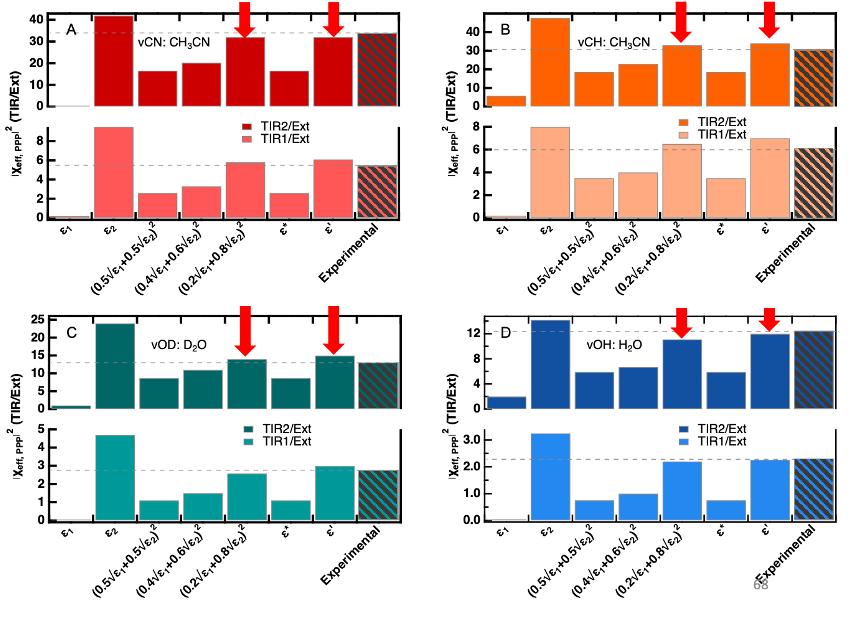}% Here is how to import EPS art
\caption{\label{fig3:wide} The ratio of $\big|\chi_{\mathrm{eff},\; PPP}^{(2)}\big|^2$ of the total internal reflection versus the external reflection geometry for $\left.\mathrm{A}\right) \mathrm{C}\equiv\mathrm{N}$ stretch of $\left.\mathrm{CH}_3 \mathrm{CN}, \mathrm{B}\right) \mathrm{CH}$ stretch of $\left.\mathrm{CH}_3 \mathrm{CN}, \mathrm{C}\right) \mathrm{OD}$ stretch of $\mathrm{D}_2 \mathrm{O}$ and, D) the $\mathrm{OH}$ stretch of $\mathrm{H}_2 \mathrm{O}$ at the $\alpha-\mathrm{Al}_2 \mathrm{O}_3$ interfaces. Our expression and six different models for $\varepsilon^{\prime}$ including: $\varepsilon_1, \varepsilon_2,\left(0.5 \sqrt{\varepsilon_1}+0.5 \sqrt{\varepsilon_2}\right)^2,\left(0.4 \sqrt{\varepsilon_1}+0.6 \sqrt{\varepsilon_2}\right)^2$, $\left(0.2 \sqrt{\varepsilon_1}+0.8 \sqrt{\varepsilon_2}\right)^2, \varepsilon^*=\varepsilon_2\left(\varepsilon_2+5\right) /\left(4 \varepsilon_2+2\right)$ and $\varepsilon^{\prime}=\varepsilon_1 \varepsilon_2\left(\varepsilon_2-\varepsilon_1+6\right) / 2\left(2 \varepsilon_2+\varepsilon_1\right)$ compared to the experimental results (last bar on the right). Gray dashed line indicates the experimental value. The top row in each panel refers to TIR2 and the bottom row refers to TIR1.}
\end{figure*}

\begin{figure*}
\includegraphics{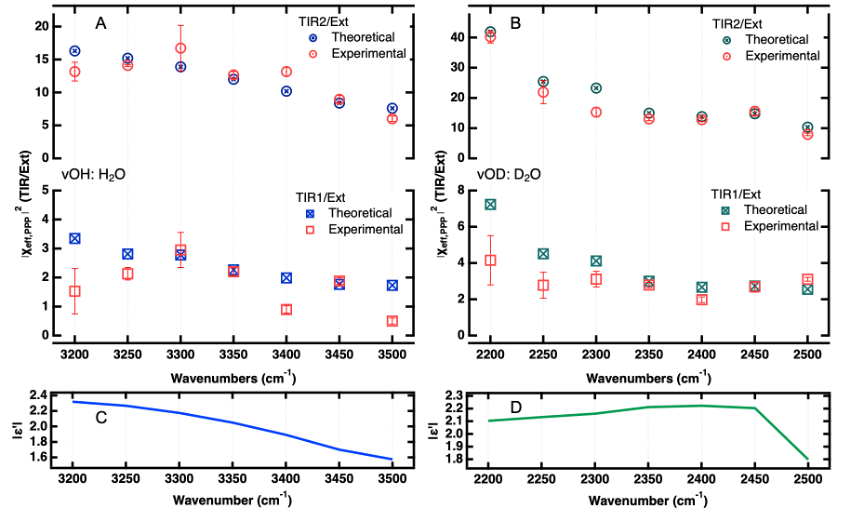}% Here is how to import EPS art
\caption{\label{fig4:wide} Calculated frequency dependent $\big|\chi_{\mathrm{eff},\; PPP}^{(2)}\big|^2$ ratio for TIR2 (Top) and TIR1 (Bottom) with respect to the external geometry for the A) $\mathrm{OH}$ stretch of $\mathrm{H}_2 \mathrm{O}$ (blue) B) OD stretch of $\mathrm{D}_2 \mathrm{O}$ (green) at the $\alpha{-}\mathrm{Al}_2\mathrm{O}_3$ interfaces compared to the experimental data (red). The dependence of $\left|\varepsilon^{\prime}\right|$ at the C) $\mathrm{H}_2 \mathrm{O} / \alpha{-}\mathrm{Al}_2\mathrm{O}_3$ and the $\left.\mathrm{D}\right) \mathrm{D}_2 \mathrm{O} / \alpha{-}\mathrm{Al}_2\mathrm{O}_3$ interfaces. The expression $\varepsilon^{\prime}=\varepsilon_1 \varepsilon_2\left(\varepsilon_2-\varepsilon_1+6\right) / 2\left(2 \varepsilon_2+\varepsilon_1\right)$, was used to calculate the interfacial dielectric constant.}
\end{figure*}

To evaluate $\varepsilon'$ we compared the experimental ratios of the effective second order nonlinear susceptibility for the PPP polarization combination $\big(\big|\chi_{\mathrm{eff},\; PPP}^{(2)}\big|^2\big)$ for TIR2 and TIR1 in relation to the external geometry (FIG. \ref{fig3:wide}) with the calculated values for different spectral regions. These include the $\mathrm{C} \equiv \mathrm{N}$ stretch (FIG. \ref{fig3:wide}A) and the $\mathrm{CH}$ stretch (FIG \ref{fig3:wide}B) of $\mathrm{CH}_3 \mathrm{CN}$, the OD stretch of $\mathrm{D}_2 \mathrm{O}$ at $2350 \mathrm{~cm}^{-1}$ (FIG \ref{fig3:wide}C) and the $\mathrm{OH}$ stretch of $\mathrm{H}_2 \mathrm{O}$ at $3350 \mathrm{~cm}^{-1}$ (FIG. \ref{fig3:wide}D). Utilizing our expression and six other models for $\varepsilon^{\prime}$ (Table S2, Supplemental Material section VI), we calculated the $\big|\chi_{\mathrm{eff},\; PPP}^{(2)}\big|^2$ ratio for each geometry using Eq.~\ref{subeq:4} and the Fresnel coefficients of $L_{X X}$ $\left(\omega_I\right), L_{Y Y}\left(\omega_I\right)$ and $L_{Z Z}\left(\omega_I\right)$ described in  Eq.~\ref{subeq:5a} - Eq.~\ref{subeq:5c}.
The $\big|\chi_{\mathrm{eff},\; PPP}^{(2)}\big|^2$ ratio comparison shows that our expression $\varepsilon_1 \varepsilon_2\left(\varepsilon_2-\varepsilon_1+6\right) / 2\left(2 \varepsilon_2+\right.$ $\varepsilon_1$ ) and the expression from MD simulations \cite{wang2017}, marked by red arrows in FIG. \ref{fig3:wide}, estimate $\varepsilon^{\prime}$ better than other models for a broad range of spectral regions investigated here. By comparing the calculated $\big|\chi_{\mathrm{eff},\; PPP}^{(2)}\big|^2$ ratio with the experimental values, we showed that all other models underestimated $\varepsilon^{\prime}$ by factors of 5 to 10 compared to the experiments. (FIG. S2: Supplemental Material section VI). Additionally, the model using $\varepsilon_2$ as $\varepsilon^{\prime}$, i.e., only considering the dielectric constant of the bulk liquid phase, overestimates the interfacial dielectric constant and ignores the impact of the solid on $\varepsilon'$. The formula $\varepsilon^*=\varepsilon_2\left(\varepsilon_2+5\right) /\left(4 \varepsilon_2+2\right)$ appears to effectively estimate the interfacial dielectric constant at the air/liquid and air/solid interface \cite{chiang2022,zhuang1999,shiratori2011}. However, our experimental results indicate that $\varepsilon^*=\varepsilon_2\left(\varepsilon_2+5\right) /\left(4 \varepsilon_2+2\right)$ might not be accurate for estimating $\varepsilon^{\prime}$ at solid/liquid interfaces because, as shown in FIG. \ref{fig3:wide} A-D, the $\big|\chi_{\mathrm{eff},\; PPP}^{(2)}\big|^2$ ratio calculated using this formula is lower than the experimental ratio. This is not surprising since the dielectric constant of the solid is different from that of air.\\
\indent The vSFG response of water at various mineral interfaces reflects the differences in the hydrogen bonding environment of the first few layers \cite{woutersen1998}. This diversity in interactions becomes evident when probing the $\mathrm{OH}$ and $\mathrm{OD}$ stretch region, \cite{woutersen1998}. It is important to note that the Fresnel factors exhibit a noticeable frequency dependence, primarily driven by the frequency-dependent dielectric constant of bulk water in the infrared region \cite{backus2012}.\\
\indent To further investigate if our expression, $\varepsilon^{\prime}=\varepsilon_1 \varepsilon_2\left(\varepsilon_2-\varepsilon_1+6\right) / 2\left(2 \varepsilon_2+\varepsilon_1\right)$, for the interfacial dielectric constant provides consistent results with respect to frequency, we compared the $\big|\chi_{\mathrm{eff},\; PPP}^{(2)}\big|^2$ for TIR2 and TIR1 versus the external geometry at different frequencies in the $\mathrm{OH}$ and OD stretch regions. The results indicate that our expression provides accurate estimates of the modulus of the interfacial dielectric constant for these frequencies (FIG. \ref{fig4:wide}: A and B). Between the six different expressions, only our expression and the one from classical MD simulations \cite{wang2017}, $\varepsilon^{\prime}$ $=\left(0.2 \sqrt{\varepsilon_1}+0.8 \sqrt{\varepsilon_2}\right)^2$, show good agreement with the experimental results at the $\mathrm{OH}$ (FIG. \ref{fig3:wide}A) and $\mathrm{OD}$ (FIG. \ref{fig3:wide}B) stretch frequencies (FIG. S9: A and B).  Our expression, derived from a continuum electrostatic approach, agrees well with the microscopic approach intrinsic to MD simulations further validates our model.  \setlength{\textfloatsep}{10pt plus 1.0pt minus 2.0pt}
\indent Using our expression for the interfacial dielectric constant for both $\mathrm{OH}$ of $\mathrm{H}_2 \mathrm{O}$ and $\mathrm{OD}$ stretching of $\mathrm{D}_2 \mathrm{O}$ reveals that $\left|\varepsilon^{\prime}\right|$ shows frequency dependency, as expected, and and varies between 1.6-2.4 (FIG. \ref{fig4:wide}C) and 1.8-2.4 (FIG. \ref{fig4:wide}D) at these interfaces, respectively. In addition, $\left|\varepsilon^{\prime}\right|$ for $\mathrm{H}_2 \mathrm{O}$ and $\mathrm{D}_2 \mathrm{O}$ at the interface appears higher than their bulk values, which could arise from the influence of $\mathrm{Al}_2 \mathrm{O}_3$ surface \cite{laurent1983}. Our value for the interfacial dielectric constant of $\mathrm{H}_2 \mathrm{O}$ and $\mathrm{D}_2 \mathrm{O}$, calculated from our expression and corroborated by results from MD simulations (FIG. S9C and D), shows that $\left|\varepsilon^{\prime}\right|$ is close to the value of the static dielectric constant $(\varepsilon \sim 2)$ determined under confinement \cite{fumagalli2018}. Additionally, using our expression we can estimate that $\left|\varepsilon^{\prime}\right|$ for the $\mathrm{C}\!\equiv\!\mathrm{N}$ and the $\mathrm{CH}$ stretch of $\mathrm{CH}_3 \mathrm{CN}$ are $\sim 2.28$ and 2.04, respectively, are higher compared to their bulk values, $\sim 1.69$ and 1.78, respectively due to the influence of $\mathrm{Al}_2 \mathrm{O}_3$ surface \cite{moutzouris2014},
\\ \indent
\textit{Conclusions.---}We have derived a generalized expression for the interfacial dielectric constant at buried, solid/liquid interfaces, $\varepsilon^{\prime}=\varepsilon_1 \varepsilon_2\left(\varepsilon_2-\varepsilon_1+6\right) / 2\left(2 \varepsilon_2+\varepsilon_1\right)$, using a slab model for a half-solvated sphere, and employed angle-dependent vSFG spectroscopy and Fresnel factor calculations to evaluate this expression in different infrared spectral regions. We showed that our approach is effective in predicting the interfacial dielectric constant across a broad frequency range for several liquids $\left(\mathrm{H}_2 \mathrm{O}, \mathrm{D}_2 \mathrm{O}\right.$ and acetonitrile) in contact with $\mathrm{Al}_2 \mathrm{O}_3$ surfaces. We expect that our expression will be valid across various other types of buried interfaces which is fundamentally important for heterogeneous chemical reactions like hydrolysis, hydrogenation, and crosscoupling reactions, etc.
\\ \indent
\textit{Acknowledgements.---}EB thanks the National Science Foundation (CHE 2102557) for supporting this work.  N.A., Z.W., and K.M.R. were supported by the U.S. Department of Energy (DOE), Office of Science, Office of Basic Energy Sciences, Chemical Sciences, Geosciences and Biosciences Division through its Geosciences program at PNNL (FWP \#56674). A portion of the research was performed using EMSL, a DOE Office of Science User Facility sponsored by the Office of Biological and Environmental Research. PNNL is a multi-program national laboratory operated for the DOE by Battelle Memorial Institute under Contract No. DE-AC05-76RL01830. 
\nocite{*}

\bibliography{apssamp}% Produces the bibliography via BibTeX.

\providecommand{\noopsort}[1]{}\providecommand{\singleletter}[1]{#1}%
\begin{thebibliography}{49}%
\makeatletter
\providecommand \@ifxundefined [1]{%
 \@ifx{#1\undefined}
}%
\providecommand \@ifnum [1]{%
 \ifnum #1\expandafter \@firstoftwo
 \else \expandafter \@secondoftwo
 \fi
}%
\providecommand \@ifx [1]{%
 \ifx #1\expandafter \@firstoftwo
 \else \expandafter \@secondoftwo
 \fi
}%
\providecommand \natexlab [1]{#1}%
\providecommand \enquote  [1]{``#1''}%
\providecommand \bibnamefont  [1]{#1}%
\providecommand \bibfnamefont [1]{#1}%
\providecommand \citenamefont [1]{#1}%
\providecommand \href@noop [0]{\@secondoftwo}%
\providecommand \href [0]{\begingroup \@sanitize@url \@href}%
\providecommand \@href[1]{\@@startlink{#1}\@@href}%
\providecommand \@@href[1]{\endgroup#1\@@endlink}%
\providecommand \@sanitize@url [0]{\catcode `\\12\catcode `\$12\catcode `\&12\catcode `\#12\catcode `\^12\catcode `\_12\catcode `\%12\relax}%
\providecommand \@@startlink[1]{}%
\providecommand \@@endlink[0]{}%
\providecommand \url  [0]{\begingroup\@sanitize@url \@url }%
\providecommand \@url [1]{\endgroup\@href {#1}{\urlprefix }}%
\providecommand \urlprefix  [0]{URL }%
\providecommand \Eprint [0]{\href }%
\providecommand \doibase [0]{https://doi.org/}%
\providecommand \selectlanguage [0]{\@gobble}%
\providecommand \bibinfo  [0]{\@secondoftwo}%
\providecommand \bibfield  [0]{\@secondoftwo}%
\providecommand \translation [1]{[#1]}%
\providecommand \BibitemOpen [0]{}%
\providecommand \bibitemStop [0]{}%
\providecommand \bibitemNoStop [0]{.\EOS\space}%
\providecommand \EOS [0]{\spacefactor3000\relax}%
\providecommand \BibitemShut  [1]{\csname bibitem#1\endcsname}%
\let\auto@bib@innerbib\@empty
%</preamble>
\bibitem [{\citenamefont {Sit}(2013)}]{sit2013}%
  \BibitemOpen
  \bibfield  {author} {\bibinfo {author} {\bibfnamefont {P.}~\bibnamefont {Sit}},\ }\href@noop {} {\emph {\bibinfo {title} {Characterization of Biomaterials}}}\ (\bibinfo  {publisher} {Woodhead Publishing},\ \bibinfo {year} {2013})\BibitemShut {NoStop}%
\bibitem [{\citenamefont {Eisenthal}(1996)}]{eisenthal1996}%
  \BibitemOpen
  \bibfield  {author} {\bibinfo {author} {\bibfnamefont {K.}~\bibnamefont {Eisenthal}},\ }\bibfield  {title} {\bibinfo {title} {Liquid interfaces probed by second-harmonic and sum-frequency spectroscopy},\ }\href@noop {} {\bibfield  {journal} {\bibinfo  {journal} {Chemical Reviews}\ }\textbf {\bibinfo {volume} {96}},\ \bibinfo {pages} {1343} (\bibinfo {year} {1996})}\BibitemShut {NoStop}%
\bibitem [{\citenamefont {Fumagalli}\ \emph {et~al.}(2018)\citenamefont {Fumagalli}, \citenamefont {Esfandiar}, \citenamefont {Fabregas}, \citenamefont {Hu}, \citenamefont {Ares},\ and\ \citenamefont {et~al.}}]{fumagalli2018}%
  \BibitemOpen
  \bibfield  {author} {\bibinfo {author} {\bibfnamefont {L.}~\bibnamefont {Fumagalli}}, \bibinfo {author} {\bibfnamefont {A.}~\bibnamefont {Esfandiar}}, \bibinfo {author} {\bibfnamefont {R.}~\bibnamefont {Fabregas}}, \bibinfo {author} {\bibfnamefont {S.}~\bibnamefont {Hu}}, \bibinfo {author} {\bibfnamefont {P.}~\bibnamefont {Ares}},\ and\ \bibinfo {author} {\bibfnamefont {A.~J.}\ \bibnamefont {et~al.}},\ }\bibfield  {title} {\bibinfo {title} {Anomalously low dielectric constant of confined water},\ }\href@noop {} {\bibfield  {journal} {\bibinfo  {journal} {Science}\ }\textbf {\bibinfo {volume} {360}},\ \bibinfo {pages} {1339} (\bibinfo {year} {2018})}\BibitemShut {NoStop}%
\bibitem [{\citenamefont {Zhu}\ \emph {et~al.}(2020)\citenamefont {Zhu}, \citenamefont {Yang}, \citenamefont {Zhu}, \citenamefont {Li}, \citenamefont {He}, \citenamefont {Huang},\ and\ \citenamefont {Li}}]{zhu2020}%
  \BibitemOpen
  \bibfield  {author} {\bibinfo {author} {\bibfnamefont {H.}~\bibnamefont {Zhu}}, \bibinfo {author} {\bibfnamefont {F.}~\bibnamefont {Yang}}, \bibinfo {author} {\bibfnamefont {Y.}~\bibnamefont {Zhu}}, \bibinfo {author} {\bibfnamefont {A.}~\bibnamefont {Li}}, \bibinfo {author} {\bibfnamefont {W.}~\bibnamefont {He}}, \bibinfo {author} {\bibfnamefont {J.}~\bibnamefont {Huang}},\ and\ \bibinfo {author} {\bibfnamefont {G.}~\bibnamefont {Li}},\ }\bibfield  {title} {\bibinfo {title} {Investigation of dielectric constants of water in a nano-confined pore},\ }\href@noop {} {\bibfield  {journal} {\bibinfo  {journal} {RSC Advances}\ }\textbf {\bibinfo {volume} {10}},\ \bibinfo {pages} {8628} (\bibinfo {year} {2020})}\BibitemShut {NoStop}%
\bibitem [{\citenamefont {Wang}\ \emph {et~al.}(2019)\citenamefont {Wang}, \citenamefont {DelloStritto}, \citenamefont {Remsing}, \citenamefont {Carnevale}, \citenamefont {Klein},\ and\ \citenamefont {Borguet}}]{wang2019sodium}%
  \BibitemOpen
  \bibfield  {author} {\bibinfo {author} {\bibfnamefont {R.}~\bibnamefont {Wang}}, \bibinfo {author} {\bibfnamefont {M.}~\bibnamefont {DelloStritto}}, \bibinfo {author} {\bibfnamefont {R.~C.}\ \bibnamefont {Remsing}}, \bibinfo {author} {\bibfnamefont {V.}~\bibnamefont {Carnevale}}, \bibinfo {author} {\bibfnamefont {M.~L.}\ \bibnamefont {Klein}},\ and\ \bibinfo {author} {\bibfnamefont {E.}~\bibnamefont {Borguet}},\ }\bibfield  {title} {\bibinfo {title} {Sodium halide adsorption and water structure at the $\alpha$-alumina(0001)/water interface},\ }\href@noop {} {\bibfield  {journal} {\bibinfo  {journal} {The Journal of Physical Chemistry C}\ }\textbf {\bibinfo {volume} {123}},\ \bibinfo {pages} {15618} (\bibinfo {year} {2019})}\BibitemShut {NoStop}%
\bibitem [{\citenamefont {Boyes}\ \emph {et~al.}(2020)\citenamefont {Boyes}, \citenamefont {LaGrow}, \citenamefont {Ward}, \citenamefont {Mitchell},\ and\ \citenamefont {Gai}}]{boyes2020}%
  \BibitemOpen
  \bibfield  {author} {\bibinfo {author} {\bibfnamefont {E.~D.}\ \bibnamefont {Boyes}}, \bibinfo {author} {\bibfnamefont {A.~P.}\ \bibnamefont {LaGrow}}, \bibinfo {author} {\bibfnamefont {M.~R.}\ \bibnamefont {Ward}}, \bibinfo {author} {\bibfnamefont {R.~W.}\ \bibnamefont {Mitchell}},\ and\ \bibinfo {author} {\bibfnamefont {P.~L.}\ \bibnamefont {Gai}},\ }\bibfield  {title} {\bibinfo {title} {Single atom dynamics in chemical reactions},\ }\href@noop {} {\bibfield  {journal} {\bibinfo  {journal} {Accounts of Chemical Research}\ }\textbf {\bibinfo {volume} {53}},\ \bibinfo {pages} {390} (\bibinfo {year} {2020})}\BibitemShut {NoStop}%
\bibitem [{\citenamefont {Jimenez}\ \emph {et~al.}(2009)\citenamefont {Jimenez}, \citenamefont {Canagaratna}, \citenamefont {Donahue}, \citenamefont {Prevot}, \citenamefont {Zhang},\ and\ \citenamefont {et~al.}}]{jimenez2009}%
  \BibitemOpen
  \bibfield  {author} {\bibinfo {author} {\bibfnamefont {J.~L.}\ \bibnamefont {Jimenez}}, \bibinfo {author} {\bibfnamefont {M.}~\bibnamefont {Canagaratna}}, \bibinfo {author} {\bibfnamefont {N.}~\bibnamefont {Donahue}}, \bibinfo {author} {\bibfnamefont {A.}~\bibnamefont {Prevot}}, \bibinfo {author} {\bibfnamefont {Q.}~\bibnamefont {Zhang}},\ and\ \bibinfo {author} {\bibfnamefont {J.~H.~K.}\ \bibnamefont {et~al.}},\ }\bibfield  {title} {\bibinfo {title} {Evolution of organic aerosols in the atmosphere},\ }\href@noop {} {\bibfield  {journal} {\bibinfo  {journal} {Science}\ }\textbf {\bibinfo {volume} {326}},\ \bibinfo {pages} {1525} (\bibinfo {year} {2009})}\BibitemShut {NoStop}%
\bibitem [{\citenamefont {Yu}\ \emph {et~al.}(2021)\citenamefont {Yu}, \citenamefont {Seki}, \citenamefont {Yu}, \citenamefont {Zhong}, \citenamefont {Sun},\ and\ \citenamefont {et~al.}}]{yu2021}%
  \BibitemOpen
  \bibfield  {author} {\bibinfo {author} {\bibfnamefont {X.}~\bibnamefont {Yu}}, \bibinfo {author} {\bibfnamefont {T.}~\bibnamefont {Seki}}, \bibinfo {author} {\bibfnamefont {C.-C.}\ \bibnamefont {Yu}}, \bibinfo {author} {\bibfnamefont {K.}~\bibnamefont {Zhong}}, \bibinfo {author} {\bibfnamefont {S.}~\bibnamefont {Sun}},\ and\ \bibinfo {author} {\bibfnamefont {M.~O.}\ \bibnamefont {et~al.}},\ }\bibfield  {title} {\bibinfo {title} {Interfacial water structure of binary liquid mixtures reflects nonideal behavior},\ }\href@noop {} {\bibfield  {journal} {\bibinfo  {journal} {The Journal of Physical Chemistry B}\ }\textbf {\bibinfo {volume} {125}},\ \bibinfo {pages} {10639} (\bibinfo {year} {2021})}\BibitemShut {NoStop}%
\bibitem [{\citenamefont {Zhong}\ \emph {et~al.}(2018)\citenamefont {Zhong}, \citenamefont {Kumar}, \citenamefont {Francisco},\ and\ \citenamefont {Zeng}}]{zhong2018}%
  \BibitemOpen
  \bibfield  {author} {\bibinfo {author} {\bibfnamefont {J.}~\bibnamefont {Zhong}}, \bibinfo {author} {\bibfnamefont {M.}~\bibnamefont {Kumar}}, \bibinfo {author} {\bibfnamefont {J.~S.}\ \bibnamefont {Francisco}},\ and\ \bibinfo {author} {\bibfnamefont {X.~C.}\ \bibnamefont {Zeng}},\ }\bibfield  {title} {\bibinfo {title} {Insight into chemistry on cloud/aerosol water surfaces},\ }\href@noop {} {\bibfield  {journal} {\bibinfo  {journal} {Accounts of Chemical Research}\ }\textbf {\bibinfo {volume} {51}},\ \bibinfo {pages} {1229} (\bibinfo {year} {2018})}\BibitemShut {NoStop}%
\bibitem [{\citenamefont {Deißenbeck}\ and\ \citenamefont {Wippermann}(2023)}]{deissenbeck2023}%
  \BibitemOpen
  \bibfield  {author} {\bibinfo {author} {\bibfnamefont {F.}~\bibnamefont {Deißenbeck}}\ and\ \bibinfo {author} {\bibfnamefont {S.}~\bibnamefont {Wippermann}},\ }\bibfield  {title} {\bibinfo {title} {Dielectric properties of nanoconfined water from ab initio thermopotentiostat molecular dynamics},\ }\href@noop {} {\bibfield  {journal} {\bibinfo  {journal} {Journal of Chemical Theory and Computation}\ }\textbf {\bibinfo {volume} {19}},\ \bibinfo {pages} {1035} (\bibinfo {year} {2023})}\BibitemShut {NoStop}%
\bibitem [{\citenamefont {Bagchi}(2013)}]{bagchi2013}%
  \BibitemOpen
  \bibfield  {author} {\bibinfo {author} {\bibfnamefont {B.}~\bibnamefont {Bagchi}},\ }\href@noop {} {\emph {\bibinfo {title} {Water in biological and chemical processes: from structure and dynamics to function}}}\ (\bibinfo  {publisher} {Cambridge University Press},\ \bibinfo {year} {2013})\BibitemShut {NoStop}%
\bibitem [{\citenamefont {Lee}\ \emph {et~al.}(2019)\citenamefont {Lee}, \citenamefont {Walker}, \citenamefont {Han}, \citenamefont {Kang}, \citenamefont {Prinz}, \citenamefont {Waymouth}, \citenamefont {Nam},\ and\ \citenamefont {Zare}}]{lee2019}%
  \BibitemOpen
  \bibfield  {author} {\bibinfo {author} {\bibfnamefont {J.~K.}\ \bibnamefont {Lee}}, \bibinfo {author} {\bibfnamefont {K.~L.}\ \bibnamefont {Walker}}, \bibinfo {author} {\bibfnamefont {H.~S.}\ \bibnamefont {Han}}, \bibinfo {author} {\bibfnamefont {J.}~\bibnamefont {Kang}}, \bibinfo {author} {\bibfnamefont {F.~B.}\ \bibnamefont {Prinz}}, \bibinfo {author} {\bibfnamefont {R.~M.}\ \bibnamefont {Waymouth}}, \bibinfo {author} {\bibfnamefont {H.~G.}\ \bibnamefont {Nam}},\ and\ \bibinfo {author} {\bibfnamefont {R.~N.}\ \bibnamefont {Zare}},\ }\bibfield  {title} {\bibinfo {title} {Spontaneous generation of hydrogen peroxide from aqueous microdroplets},\ }\href@noop {} {\bibfield  {journal} {\bibinfo  {journal} {Proceedings of the National Academy of Sciences}\ }\textbf {\bibinfo {volume} {116}},\ \bibinfo {pages} {19294} (\bibinfo {year} {2019})}\BibitemShut {NoStop}%
\bibitem [{\citenamefont {Xiong}\ \emph {et~al.}(2020)\citenamefont {Xiong}, \citenamefont {Lee}, \citenamefont {Zare},\ and\ \citenamefont {Min}}]{xiong2020}%
  \BibitemOpen
  \bibfield  {author} {\bibinfo {author} {\bibfnamefont {H.}~\bibnamefont {Xiong}}, \bibinfo {author} {\bibfnamefont {J.~K.}\ \bibnamefont {Lee}}, \bibinfo {author} {\bibfnamefont {R.~N.}\ \bibnamefont {Zare}},\ and\ \bibinfo {author} {\bibfnamefont {W.}~\bibnamefont {Min}},\ }\bibfield  {title} {\bibinfo {title} {Strong electric field observed at the interface of aqueous microdroplets},\ }\href@noop {} {\bibfield  {journal} {\bibinfo  {journal} {Physical Chemistry Letters}\ }\textbf {\bibinfo {volume} {11}},\ \bibinfo {pages} {7423} (\bibinfo {year} {2020})}\BibitemShut {NoStop}%
\bibitem [{\citenamefont {Teschke}\ \emph {et~al.}(2001)\citenamefont {Teschke}, \citenamefont {Ceotto},\ and\ \citenamefont {Souza}}]{teschke2001}%
  \BibitemOpen
  \bibfield  {author} {\bibinfo {author} {\bibfnamefont {O.}~\bibnamefont {Teschke}}, \bibinfo {author} {\bibfnamefont {G.}~\bibnamefont {Ceotto}},\ and\ \bibinfo {author} {\bibfnamefont {E.~D.}\ \bibnamefont {Souza}},\ }\bibfield  {title} {\bibinfo {title} {Interfacial water dielectric-permittivity-profile measurements using atomic force microscopy},\ }\href@noop {} {\bibfield  {journal} {\bibinfo  {journal} {Physical Review E}\ }\textbf {\bibinfo {volume} {64}},\ \bibinfo {pages} {011605} (\bibinfo {year} {2001})}\BibitemShut {NoStop}%
\bibitem [{\citenamefont {Chiang}\ \emph {et~al.}(2022)\citenamefont {Chiang}, \citenamefont {Seki}, \citenamefont {Yu}, \citenamefont {Ohto}, \citenamefont {Hunger}, \citenamefont {Bonn},\ and\ \citenamefont {Nagata}}]{chiang2022}%
  \BibitemOpen
  \bibfield  {author} {\bibinfo {author} {\bibfnamefont {K.-Y.}\ \bibnamefont {Chiang}}, \bibinfo {author} {\bibfnamefont {T.}~\bibnamefont {Seki}}, \bibinfo {author} {\bibfnamefont {C.-C.}\ \bibnamefont {Yu}}, \bibinfo {author} {\bibfnamefont {T.}~\bibnamefont {Ohto}}, \bibinfo {author} {\bibfnamefont {J.}~\bibnamefont {Hunger}}, \bibinfo {author} {\bibfnamefont {M.}~\bibnamefont {Bonn}},\ and\ \bibinfo {author} {\bibfnamefont {Y.}~\bibnamefont {Nagata}},\ }\bibfield  {title} {\bibinfo {title} {The dielectric function profile across the water interface through surface-specific vibrational spectroscopy and simulations},\ }\href@noop {} {\bibfield  {journal} {\bibinfo  {journal} {Proceedings of the National Academy of Sciences}\ }\textbf {\bibinfo {volume} {119}},\ \bibinfo {pages} {e2204156119} (\bibinfo {year} {2022})}\BibitemShut {NoStop}%
\bibitem [{\citenamefont {Loche}\ \emph {et~al.}(2020)\citenamefont {Loche}, \citenamefont {Ayaz}, \citenamefont {Wolde-Kidan}, \citenamefont {Schlaich},\ and\ \citenamefont {Netz}}]{loche2020}%
  \BibitemOpen
  \bibfield  {author} {\bibinfo {author} {\bibfnamefont {P.}~\bibnamefont {Loche}}, \bibinfo {author} {\bibfnamefont {C.}~\bibnamefont {Ayaz}}, \bibinfo {author} {\bibfnamefont {A.}~\bibnamefont {Wolde-Kidan}}, \bibinfo {author} {\bibfnamefont {A.}~\bibnamefont {Schlaich}},\ and\ \bibinfo {author} {\bibfnamefont {R.~R.}\ \bibnamefont {Netz}},\ }\bibfield  {title} {\bibinfo {title} {Universal and nonuniversal aspects of electrostatics in aqueous nanoconfinement},\ }\href@noop {} {\bibfield  {journal} {\bibinfo  {journal} {The Journal of Physical Chemistry B}\ }\textbf {\bibinfo {volume} {124}},\ \bibinfo {pages} {4365} (\bibinfo {year} {2020})}\BibitemShut {NoStop}%
\bibitem [{\citenamefont {Munoz-Santiburcio}\ and\ \citenamefont {Marx}(2017)}]{munoz-santiburcio2017}%
  \BibitemOpen
  \bibfield  {author} {\bibinfo {author} {\bibfnamefont {D.}~\bibnamefont {Munoz-Santiburcio}}\ and\ \bibinfo {author} {\bibfnamefont {D.}~\bibnamefont {Marx}},\ }\bibfield  {title} {\bibinfo {title} {Chemistry in nanoconfined water},\ }\href@noop {} {\bibfield  {journal} {\bibinfo  {journal} {Chemical Science}\ }\textbf {\bibinfo {volume} {8}},\ \bibinfo {pages} {3444} (\bibinfo {year} {2017})}\BibitemShut {NoStop}%
\bibitem [{\citenamefont {Xiao}\ \emph {et~al.}(2019)\citenamefont {Xiao}, \citenamefont {Yang}, \citenamefont {Zhang}, \citenamefont {Feng}, \citenamefont {Zhang}, \citenamefont {Jiang}, \citenamefont {Sui},\ and\ \citenamefont {Jia}}]{xiao2019}%
  \BibitemOpen
  \bibfield  {author} {\bibinfo {author} {\bibfnamefont {H.}~\bibnamefont {Xiao}}, \bibinfo {author} {\bibfnamefont {Z.}~\bibnamefont {Yang}}, \bibinfo {author} {\bibfnamefont {L.}~\bibnamefont {Zhang}}, \bibinfo {author} {\bibfnamefont {D.}~\bibnamefont {Feng}}, \bibinfo {author} {\bibfnamefont {F.}~\bibnamefont {Zhang}}, \bibinfo {author} {\bibfnamefont {M.}~\bibnamefont {Jiang}}, \bibinfo {author} {\bibfnamefont {Q.}~\bibnamefont {Sui}},\ and\ \bibinfo {author} {\bibfnamefont {L.}~\bibnamefont {Jia}},\ }\bibfield  {title} {\bibinfo {title} {Compositional analysis of cement raw meal by near-infrared (nir) spectroscopy},\ }\href@noop {} {\bibfield  {journal} {\bibinfo  {journal} {Analytical Letters}\ }\textbf {\bibinfo {volume} {52}},\ \bibinfo {pages} {2931} (\bibinfo {year} {2019})}\BibitemShut {NoStop}%
\bibitem [{\citenamefont {Zhan}\ \emph {et~al.}(2020)\citenamefont {Zhan}, \citenamefont {Liu}, \citenamefont {Tian},\ and\ \citenamefont {Ren}}]{zhan2020}%
  \BibitemOpen
  \bibfield  {author} {\bibinfo {author} {\bibfnamefont {C.}~\bibnamefont {Zhan}}, \bibinfo {author} {\bibfnamefont {B.-W.}\ \bibnamefont {Liu}}, \bibinfo {author} {\bibfnamefont {Z.-Q.}\ \bibnamefont {Tian}},\ and\ \bibinfo {author} {\bibfnamefont {B.}~\bibnamefont {Ren}},\ }\bibfield  {title} {\bibinfo {title} {Determining the interfacial refractive index via ultrasensitive plasmonic sensors},\ }\href@noop {} {\bibfield  {journal} {\bibinfo  {journal} {Journal of the American Chemical Society}\ }\textbf {\bibinfo {volume} {142}},\ \bibinfo {pages} {10905} (\bibinfo {year} {2020})}\BibitemShut {NoStop}%
\bibitem [{\citenamefont {Teschke}\ and\ \citenamefont {Souza}(2005)}]{teschke2005}%
  \BibitemOpen
  \bibfield  {author} {\bibinfo {author} {\bibfnamefont {O.}~\bibnamefont {Teschke}}\ and\ \bibinfo {author} {\bibfnamefont {E.~D.}\ \bibnamefont {Souza}},\ }\bibfield  {title} {\bibinfo {title} {Water molecule clusters measured at water/air interfaces using atomic force microscopy},\ }\href@noop {} {\bibfield  {journal} {\bibinfo  {journal} {Physical Chemistry Chemical Physics}\ }\textbf {\bibinfo {volume} {7}},\ \bibinfo {pages} {3856} (\bibinfo {year} {2005})}\BibitemShut {NoStop}%
\bibitem [{\citenamefont {Underwood}\ and\ \citenamefont {Bourg}(2022)}]{underwood2022}%
  \BibitemOpen
  \bibfield  {author} {\bibinfo {author} {\bibfnamefont {T.~R.}\ \bibnamefont {Underwood}}\ and\ \bibinfo {author} {\bibfnamefont {I.~C.}\ \bibnamefont {Bourg}},\ }\bibfield  {title} {\bibinfo {title} {Dielectric properties of water in charged nanopores},\ }\href@noop {} {\bibfield  {journal} {\bibinfo  {journal} {The Journal of Physical Chemistry B}\ }\textbf {\bibinfo {volume} {126}},\ \bibinfo {pages} {2688} (\bibinfo {year} {2022})}\BibitemShut {NoStop}%
\bibitem [{\citenamefont {Othman}\ \emph {et~al.}(2011)\citenamefont {Othman}, \citenamefont {Ramli}, \citenamefont {Tyng}, \citenamefont {Ahmad},\ and\ \citenamefont {Akil}}]{othman2011}%
  \BibitemOpen
  \bibfield  {author} {\bibinfo {author} {\bibfnamefont {M.~B.~H.}\ \bibnamefont {Othman}}, \bibinfo {author} {\bibfnamefont {M.~R.}\ \bibnamefont {Ramli}}, \bibinfo {author} {\bibfnamefont {L.~Y.}\ \bibnamefont {Tyng}}, \bibinfo {author} {\bibfnamefont {Z.}~\bibnamefont {Ahmad}},\ and\ \bibinfo {author} {\bibfnamefont {H.~M.}\ \bibnamefont {Akil}},\ }\bibfield  {title} {\bibinfo {title} {Dielectric constant and refractive index of poly (siloxane–imide) block copolymer},\ }\href@noop {} {\bibfield  {journal} {\bibinfo  {journal} {Materials \& Design}\ }\textbf {\bibinfo {volume} {32}},\ \bibinfo {pages} {3173} (\bibinfo {year} {2011})}\BibitemShut {NoStop}%
\bibitem [{\citenamefont {Shen}(1989)}]{shen1989}%
  \BibitemOpen
  \bibfield  {author} {\bibinfo {author} {\bibfnamefont {Y.}~\bibnamefont {Shen}},\ }\bibfield  {title} {\bibinfo {title} {Surface properties probed by second-harmonic and sum-frequency generation},\ }\href@noop {} {\bibfield  {journal} {\bibinfo  {journal} {Nature}\ }\textbf {\bibinfo {volume} {337}},\ \bibinfo {pages} {519} (\bibinfo {year} {1989})}\BibitemShut {NoStop}%
\bibitem [{\citenamefont {Wang}\ \emph {et~al.}(2021)\citenamefont {Wang}, \citenamefont {Klein}, \citenamefont {Carnevale},\ and\ \citenamefont {Borguet}}]{wang2021}%
  \BibitemOpen
  \bibfield  {author} {\bibinfo {author} {\bibfnamefont {R.}~\bibnamefont {Wang}}, \bibinfo {author} {\bibfnamefont {M.~L.}\ \bibnamefont {Klein}}, \bibinfo {author} {\bibfnamefont {V.}~\bibnamefont {Carnevale}},\ and\ \bibinfo {author} {\bibfnamefont {E.}~\bibnamefont {Borguet}},\ }\bibfield  {title} {\bibinfo {title} {Investigations of water/oxide interfaces by molecular dynamics simulations},\ }\href@noop {} {\bibfield  {journal} {\bibinfo  {journal} {Wiley Interdisciplinary Reviews: Computational Molecular Science}\ }\textbf {\bibinfo {volume} {11}},\ \bibinfo {pages} {e1537} (\bibinfo {year} {2021})}\BibitemShut {NoStop}%
\bibitem [{\citenamefont {Morita}(2018)}]{morita2018}%
  \BibitemOpen
  \bibfield  {author} {\bibinfo {author} {\bibfnamefont {A.}~\bibnamefont {Morita}},\ }\href@noop {} {\emph {\bibinfo {title} {Theory of sum frequency generation spectroscopy}}},\ Lecture Notes in Chemistry\ (\bibinfo  {publisher} {Springer},\ \bibinfo {year} {2018})\BibitemShut {NoStop}%
\bibitem [{\citenamefont {Zhuang}\ \emph {et~al.}(1999)\citenamefont {Zhuang}, \citenamefont {Miranda}, \citenamefont {Kim},\ and\ \citenamefont {Shen}}]{zhuang1999}%
  \BibitemOpen
  \bibfield  {author} {\bibinfo {author} {\bibfnamefont {X.}~\bibnamefont {Zhuang}}, \bibinfo {author} {\bibfnamefont {P.}~\bibnamefont {Miranda}}, \bibinfo {author} {\bibfnamefont {D.}~\bibnamefont {Kim}},\ and\ \bibinfo {author} {\bibfnamefont {Y.}~\bibnamefont {Shen}},\ }\bibfield  {title} {\bibinfo {title} {Mapping molecular orientation and conformation at interfaces by surface nonlinear optics},\ }\href@noop {} {\bibfield  {journal} {\bibinfo  {journal} {Physical Review B}\ }\textbf {\bibinfo {volume} {59}},\ \bibinfo {pages} {12632} (\bibinfo {year} {1999})}\BibitemShut {NoStop}%
\bibitem [{\citenamefont {Wang}\ \emph {et~al.}(2017)\citenamefont {Wang}, \citenamefont {Ishiyama},\ and\ \citenamefont {Morita}}]{wang2017}%
  \BibitemOpen
  \bibfield  {author} {\bibinfo {author} {\bibfnamefont {L.}~\bibnamefont {Wang}}, \bibinfo {author} {\bibfnamefont {T.}~\bibnamefont {Ishiyama}},\ and\ \bibinfo {author} {\bibfnamefont {A.}~\bibnamefont {Morita}},\ }\bibfield  {title} {\bibinfo {title} {Theoretical investigation of $\text{C}-\text{H}$ vibrational spectroscopy. 1. modeling of methyl and methylene groups of ethanol with different conformers},\ }\href@noop {} {\bibfield  {journal} {\bibinfo  {journal} {The Journal of Physical Chemistry A}\ }\textbf {\bibinfo {volume} {121}},\ \bibinfo {pages} {6687} (\bibinfo {year} {2017})}\BibitemShut {NoStop}%
\bibitem [{\citenamefont {Shiratori}\ and\ \citenamefont {Morita}(2011)}]{shiratori2011}%
  \BibitemOpen
  \bibfield  {author} {\bibinfo {author} {\bibfnamefont {K.}~\bibnamefont {Shiratori}}\ and\ \bibinfo {author} {\bibfnamefont {A.}~\bibnamefont {Morita}},\ }\bibfield  {title} {\bibinfo {title} {Molecular theory on dielectric constant at interfaces: A molecular dynamics study of the water/vapor interface},\ }\href@noop {} {\bibfield  {journal} {\bibinfo  {journal} {The Journal of Chemical Physics}\ }\textbf {\bibinfo {volume} {134}},\ \bibinfo {pages} {234705} (\bibinfo {year} {2011})}\BibitemShut {NoStop}%
\bibitem [{\citenamefont {Malitson}(1962)}]{malitson1962}%
  \BibitemOpen
  \bibfield  {author} {\bibinfo {author} {\bibfnamefont {I.~H.}\ \bibnamefont {Malitson}},\ }\bibfield  {title} {\bibinfo {title} {Refraction and dispersion of synthetic sapphire},\ }\href@noop {} {\bibfield  {journal} {\bibinfo  {journal} {Journal of the Optical Society of America}\ }\textbf {\bibinfo {volume} {52}},\ \bibinfo {pages} {1377} (\bibinfo {year} {1962})}\BibitemShut {NoStop}%
\bibitem [{\citenamefont {Böttcher}(1973)}]{bottcher1973}%
  \BibitemOpen
  \bibfield  {author} {\bibinfo {author} {\bibfnamefont {C.}~\bibnamefont {Böttcher}},\ }\href@noop {} {\emph {\bibinfo {title} {Theory of Electric Polarization}}}\ (\bibinfo  {publisher} {Elsevier},\ \bibinfo {year} {1973})\BibitemShut {NoStop}%
\bibitem [{\citenamefont {Eftekhari-Bafrooei}\ and\ \citenamefont {Borguet}(2010)}]{eftekhari-bafrooei2010}%
  \BibitemOpen
  \bibfield  {author} {\bibinfo {author} {\bibfnamefont {A.}~\bibnamefont {Eftekhari-Bafrooei}}\ and\ \bibinfo {author} {\bibfnamefont {E.}~\bibnamefont {Borguet}},\ }\bibfield  {title} {\bibinfo {title} {Effect of hydrogen-bond strength on the vibrational relaxation of interfacial water},\ }\href@noop {} {\bibfield  {journal} {\bibinfo  {journal} {Journal of the American Chemical Society}\ }\textbf {\bibinfo {volume} {132}},\ \bibinfo {pages} {3756} (\bibinfo {year} {2010})}\BibitemShut {NoStop}%
\bibitem [{\citenamefont {Lesnicki}\ \emph {et~al.}(2020)\citenamefont {Lesnicki}, \citenamefont {Zhang}, \citenamefont {Bonn}, \citenamefont {Sulpizi},\ and\ \citenamefont {Backus}}]{lesnicki2020}%
  \BibitemOpen
  \bibfield  {author} {\bibinfo {author} {\bibfnamefont {D.}~\bibnamefont {Lesnicki}}, \bibinfo {author} {\bibfnamefont {Z.}~\bibnamefont {Zhang}}, \bibinfo {author} {\bibfnamefont {M.}~\bibnamefont {Bonn}}, \bibinfo {author} {\bibfnamefont {M.}~\bibnamefont {Sulpizi}},\ and\ \bibinfo {author} {\bibfnamefont {E.~H.~G.}\ \bibnamefont {Backus}},\ }\bibfield  {title} {\bibinfo {title} {Surface charges at the $\mathrm{CaF_2}$/water interface allow very fast intermolecular vibrational-energy transfer},\ }\href@noop {} {\bibfield  {journal} {\bibinfo  {journal} {Angewandte Chemie International Edition}\ }\textbf {\bibinfo {volume} {59}},\ \bibinfo {pages} {13116} (\bibinfo {year} {2020})}\BibitemShut {NoStop}%
\bibitem [{\citenamefont {Piontek}\ and\ \citenamefont {Borguet}(2023)}]{piontek2023}%
  \BibitemOpen
  \bibfield  {author} {\bibinfo {author} {\bibfnamefont {S.~M.}\ \bibnamefont {Piontek}}\ and\ \bibinfo {author} {\bibfnamefont {E.}~\bibnamefont {Borguet}},\ }\bibfield  {title} {\bibinfo {title} {Vibrational spectroscopy of geochemical interfaces},\ }\href@noop {} {\bibfield  {journal} {\bibinfo  {journal} {Surface Science Reports}\ }\textbf {\bibinfo {volume} {78}},\ \bibinfo {pages} {100606} (\bibinfo {year} {2023})}\BibitemShut {NoStop}%
\bibitem [{\citenamefont {Wang}\ \emph {et~al.}(2015)\citenamefont {Wang}, \citenamefont {Velarde}, \citenamefont {Gan},\ and\ \citenamefont {Fu}}]{wang2015}%
  \BibitemOpen
  \bibfield  {author} {\bibinfo {author} {\bibfnamefont {H.-F.}\ \bibnamefont {Wang}}, \bibinfo {author} {\bibfnamefont {L.}~\bibnamefont {Velarde}}, \bibinfo {author} {\bibfnamefont {W.}~\bibnamefont {Gan}},\ and\ \bibinfo {author} {\bibfnamefont {L.}~\bibnamefont {Fu}},\ }\bibfield  {title} {\bibinfo {title} {Quantitative sum-frequency generation vibrational spectroscopy of molecular surfaces and interfaces: lineshape, polarization, and orientation},\ }\href@noop {} {\bibfield  {journal} {\bibinfo  {journal} {Annual Review of Physical Chemistry}\ }\textbf {\bibinfo {volume} {66}},\ \bibinfo {pages} {189} (\bibinfo {year} {2015})}\BibitemShut {NoStop}%
\bibitem [{\citenamefont {Busson}(2023)}]{busson2023sum}%
  \BibitemOpen
  \bibfield  {author} {\bibinfo {author} {\bibfnamefont {B.}~\bibnamefont {Busson}},\ }\bibfield  {title} {\bibinfo {title} {Sum-frequency generation at interfaces: A fresnel story. ii. analytical expressions for multilayer systems},\ }\href@noop {} {\bibfield  {journal} {\bibinfo  {journal} {The Journal of Chemical Physics}\ }\textbf {\bibinfo {volume} {159}} (\bibinfo {year} {2023})}\BibitemShut {NoStop}%
\bibitem [{\citenamefont {Löbau}\ and\ \citenamefont {Wolfrum}(1997)}]{lobau1997}%
  \BibitemOpen
  \bibfield  {author} {\bibinfo {author} {\bibfnamefont {J.}~\bibnamefont {Löbau}}\ and\ \bibinfo {author} {\bibfnamefont {K.}~\bibnamefont {Wolfrum}},\ }\bibfield  {title} {\bibinfo {title} {Sum-frequency spectroscopy in total internal reflection geometry: signal enhancement and access to molecular properties},\ }\href@noop {} {\bibfield  {journal} {\bibinfo  {journal} {Journal of the Optical Society of America B}\ }\textbf {\bibinfo {volume} {14}},\ \bibinfo {pages} {2505} (\bibinfo {year} {1997})}\BibitemShut {NoStop}%
\bibitem [{\citenamefont {Ba{\~n}uelos}\ \emph {et~al.}(2023)\citenamefont {Ba{\~n}uelos}, \citenamefont {Borguet}, \citenamefont {Brown~Jr}, \citenamefont {Cygan}, \citenamefont {DeYoreo}, \citenamefont {Dove}, \citenamefont {Gaigeot}, \citenamefont {Geiger}, \citenamefont {Gibbs}, \citenamefont {Grassian} \emph {et~al.}}]{banuelos2023}%
  \BibitemOpen
  \bibfield  {author} {\bibinfo {author} {\bibfnamefont {J.~L.}\ \bibnamefont {Ba{\~n}uelos}}, \bibinfo {author} {\bibfnamefont {E.}~\bibnamefont {Borguet}}, \bibinfo {author} {\bibfnamefont {G.~E.}\ \bibnamefont {Brown~Jr}}, \bibinfo {author} {\bibfnamefont {R.~T.}\ \bibnamefont {Cygan}}, \bibinfo {author} {\bibfnamefont {J.~J.}\ \bibnamefont {DeYoreo}}, \bibinfo {author} {\bibfnamefont {P.~M.}\ \bibnamefont {Dove}}, \bibinfo {author} {\bibfnamefont {M.-P.}\ \bibnamefont {Gaigeot}}, \bibinfo {author} {\bibfnamefont {F.~M.}\ \bibnamefont {Geiger}}, \bibinfo {author} {\bibfnamefont {J.~M.}\ \bibnamefont {Gibbs}}, \bibinfo {author} {\bibfnamefont {V.~H.}\ \bibnamefont {Grassian}}, \emph {et~al.},\ }\bibfield  {title} {\bibinfo {title} {Oxide–and silicate–water interfaces and their roles in technology and the environment},\ }\href@noop {} {\bibfield  {journal} {\bibinfo  {journal} {Chemical Reviews}\ }\textbf {\bibinfo {volume} {123}},\ \bibinfo {pages} {6413} (\bibinfo {year} {2023})}\BibitemShut {NoStop}%
\bibitem [{\citenamefont {Lock}\ and\ \citenamefont {Bakker}(2002)}]{lock2002}%
  \BibitemOpen
  \bibfield  {author} {\bibinfo {author} {\bibfnamefont {A.}~\bibnamefont {Lock}}\ and\ \bibinfo {author} {\bibfnamefont {H.}~\bibnamefont {Bakker}},\ }\bibfield  {title} {\bibinfo {title} {Temperature dependence of vibrational relaxation in liquid $\mathrm{H_2O}$},\ }\href@noop {} {\bibfield  {journal} {\bibinfo  {journal} {The Journal of Chemical Physics}\ }\textbf {\bibinfo {volume} {117}},\ \bibinfo {pages} {1708} (\bibinfo {year} {2002})}\BibitemShut {NoStop}%
\bibitem [{\citenamefont {Rehl}\ \emph {et~al.}(2018)\citenamefont {Rehl}, \citenamefont {Li},\ and\ \citenamefont {Gibbs}}]{rehl2018}%
  \BibitemOpen
  \bibfield  {author} {\bibinfo {author} {\bibfnamefont {B.}~\bibnamefont {Rehl}}, \bibinfo {author} {\bibfnamefont {Z.}~\bibnamefont {Li}},\ and\ \bibinfo {author} {\bibfnamefont {J.~M.}\ \bibnamefont {Gibbs}},\ }\bibfield  {title} {\bibinfo {title} {Influence of high $\mathrm{pH}$ on the organization of acetonitrile at the silica/water interface studied by sum frequency generation spectroscopy},\ }\href@noop {} {\bibfield  {journal} {\bibinfo  {journal} {Langmuir}\ }\textbf {\bibinfo {volume} {34}},\ \bibinfo {pages} {4445} (\bibinfo {year} {2018})}\BibitemShut {NoStop}%
\bibitem [{\citenamefont {Rivera}\ \emph {et~al.}(2013)\citenamefont {Rivera}, \citenamefont {Bender}, \citenamefont {Manfred},\ and\ \citenamefont {Fourkas}}]{rivera2013}%
  \BibitemOpen
  \bibfield  {author} {\bibinfo {author} {\bibfnamefont {C.~A.}\ \bibnamefont {Rivera}}, \bibinfo {author} {\bibfnamefont {J.~S.}\ \bibnamefont {Bender}}, \bibinfo {author} {\bibfnamefont {K.}~\bibnamefont {Manfred}},\ and\ \bibinfo {author} {\bibfnamefont {J.~T.}\ \bibnamefont {Fourkas}},\ }\bibfield  {title} {\bibinfo {title} {Persistence of acetonitrile bilayers at the interface of acetonitrile/water mixtures with silica},\ }\href@noop {} {\bibfield  {journal} {\bibinfo  {journal} {The Journal of Physical Chemistry A}\ }\textbf {\bibinfo {volume} {117}},\ \bibinfo {pages} {12060} (\bibinfo {year} {2013})}\BibitemShut {NoStop}%
\bibitem [{\citenamefont {Wang}\ \emph {et~al.}(2014)\citenamefont {Wang}, \citenamefont {Fried}, \citenamefont {Boxer},\ and\ \citenamefont {Markland}}]{wang2014}%
  \BibitemOpen
  \bibfield  {author} {\bibinfo {author} {\bibfnamefont {L.}~\bibnamefont {Wang}}, \bibinfo {author} {\bibfnamefont {S.~D.}\ \bibnamefont {Fried}}, \bibinfo {author} {\bibfnamefont {S.~G.}\ \bibnamefont {Boxer}},\ and\ \bibinfo {author} {\bibfnamefont {T.~E.}\ \bibnamefont {Markland}},\ }\bibfield  {title} {\bibinfo {title} {Quantum delocalization of protons in the hydrogen-bond network of an enzyme active site},\ }\href@noop {} {\bibfield  {journal} {\bibinfo  {journal} {Proceedings of the National Academy of Sciences}\ }\textbf {\bibinfo {volume} {111}},\ \bibinfo {pages} {18454} (\bibinfo {year} {2014})}\BibitemShut {NoStop}%
\bibitem [{\citenamefont {Zhang}\ \emph {et~al.}(1993)\citenamefont {Zhang}, \citenamefont {Gutow}, \citenamefont {Eisenthal},\ and\ \citenamefont {Heinz}}]{zhang1993}%
  \BibitemOpen
  \bibfield  {author} {\bibinfo {author} {\bibfnamefont {D.}~\bibnamefont {Zhang}}, \bibinfo {author} {\bibfnamefont {J.}~\bibnamefont {Gutow}}, \bibinfo {author} {\bibfnamefont {K.}~\bibnamefont {Eisenthal}},\ and\ \bibinfo {author} {\bibfnamefont {T.}~\bibnamefont {Heinz}},\ }\bibfield  {title} {\bibinfo {title} {Sudden structural change at an air/binary liquid interface: Sum frequency study of the air/acetonitrile–water interface},\ }\href@noop {} {\bibfield  {journal} {\bibinfo  {journal} {The Journal of Chemical Physics}\ }\textbf {\bibinfo {volume} {98}},\ \bibinfo {pages} {5099} (\bibinfo {year} {1993})}\BibitemShut {NoStop}%
\bibitem [{\citenamefont {Dadashi}\ \emph {et~al.}(2024)\citenamefont {Dadashi}, \citenamefont {Parshotam}, \citenamefont {Mandal}, \citenamefont {Rehl}, \citenamefont {Gibbs},\ and\ \citenamefont {Borguet}}]{dadashi2024}%
  \BibitemOpen
  \bibfield  {author} {\bibinfo {author} {\bibfnamefont {S.}~\bibnamefont {Dadashi}}, \bibinfo {author} {\bibfnamefont {S.}~\bibnamefont {Parshotam}}, \bibinfo {author} {\bibfnamefont {B.}~\bibnamefont {Mandal}}, \bibinfo {author} {\bibfnamefont {B.}~\bibnamefont {Rehl}}, \bibinfo {author} {\bibfnamefont {J.~M.}\ \bibnamefont {Gibbs}},\ and\ \bibinfo {author} {\bibfnamefont {E.}~\bibnamefont {Borguet}},\ }\bibfield  {title} {\bibinfo {title} {Influence of charged site density on local electric fields and polar solvent organization at oxide interfaces},\ }\href@noop {} {\bibfield  {journal} {\bibinfo  {journal} {The Journal of Physical Chemistry C}\ }\textbf {\bibinfo {volume} {128}},\ \bibinfo {pages} {9683–9692} (\bibinfo {year} {2024})}\BibitemShut {NoStop}%
\bibitem [{\citenamefont {Dereka}\ \emph {et~al.}(2021)\citenamefont {Dereka}, \citenamefont {Lewis}, \citenamefont {Keim}, \citenamefont {Snyder},\ and\ \citenamefont {Tokmakoff}}]{dereka2021}%
  \BibitemOpen
  \bibfield  {author} {\bibinfo {author} {\bibfnamefont {B.}~\bibnamefont {Dereka}}, \bibinfo {author} {\bibfnamefont {N.~H.}\ \bibnamefont {Lewis}}, \bibinfo {author} {\bibfnamefont {J.~H.}\ \bibnamefont {Keim}}, \bibinfo {author} {\bibfnamefont {S.~A.}\ \bibnamefont {Snyder}},\ and\ \bibinfo {author} {\bibfnamefont {A.}~\bibnamefont {Tokmakoff}},\ }\bibfield  {title} {\bibinfo {title} {Characterization of acetonitrile isotopologues as vibrational probes of electrolytes},\ }\href@noop {} {\bibfield  {journal} {\bibinfo  {journal} {The Journal of Physical Chemistry B}\ }\textbf {\bibinfo {volume} {126}},\ \bibinfo {pages} {278} (\bibinfo {year} {2021})}\BibitemShut {NoStop}%
\bibitem [{\citenamefont {Woutersen}\ \emph {et~al.}(1998)\citenamefont {Woutersen}, \citenamefont {Emmerichs}, \citenamefont {Nienhuys},\ and\ \citenamefont {Bakker}}]{woutersen1998}%
  \BibitemOpen
  \bibfield  {author} {\bibinfo {author} {\bibfnamefont {S.}~\bibnamefont {Woutersen}}, \bibinfo {author} {\bibfnamefont {U.}~\bibnamefont {Emmerichs}}, \bibinfo {author} {\bibfnamefont {H.-K.}\ \bibnamefont {Nienhuys}},\ and\ \bibinfo {author} {\bibfnamefont {H.~J.}\ \bibnamefont {Bakker}},\ }\bibfield  {title} {\bibinfo {title} {Anomalous temperature dependence of vibrational lifetimes in water and ice},\ }\href@noop {} {\bibfield  {journal} {\bibinfo  {journal} {Physical Review Letters}\ }\textbf {\bibinfo {volume} {81}},\ \bibinfo {pages} {1106} (\bibinfo {year} {1998})}\BibitemShut {NoStop}%
\bibitem [{\citenamefont {Backus}\ \emph {et~al.}(2012)\citenamefont {Backus}, \citenamefont {Garcia-Araez}, \citenamefont {Bonn},\ and\ \citenamefont {Bakker}}]{backus2012}%
  \BibitemOpen
  \bibfield  {author} {\bibinfo {author} {\bibfnamefont {E.~H.}\ \bibnamefont {Backus}}, \bibinfo {author} {\bibfnamefont {N.}~\bibnamefont {Garcia-Araez}}, \bibinfo {author} {\bibfnamefont {M.}~\bibnamefont {Bonn}},\ and\ \bibinfo {author} {\bibfnamefont {H.~J.}\ \bibnamefont {Bakker}},\ }\bibfield  {title} {\bibinfo {title} {On the role of $\mathrm{Fresnel}$ factors in sum-frequency generation spectroscopy of metal–water and metal-oxide–water interfaces},\ }\href@noop {} {\bibfield  {journal} {\bibinfo  {journal} {The Journal of Physical Chemistry C}\ }\textbf {\bibinfo {volume} {116}},\ \bibinfo {pages} {23351} (\bibinfo {year} {2012})}\BibitemShut {NoStop}%
\bibitem [{\citenamefont {Laurent}\ \emph {et~al.}(1983)\citenamefont {Laurent}, \citenamefont {Billiet},\ and\ \citenamefont {Galan}}]{laurent1983}%
  \BibitemOpen
  \bibfield  {author} {\bibinfo {author} {\bibfnamefont {C.}~\bibnamefont {Laurent}}, \bibinfo {author} {\bibfnamefont {H.}~\bibnamefont {Billiet}},\ and\ \bibinfo {author} {\bibfnamefont {L.~D.}\ \bibnamefont {Galan}},\ }\bibfield  {title} {\bibinfo {title} {On the use of alumina in \uppercase{HPLC} with aqueous mobile phases at extreme $\mathrm{pH}$},\ }\href@noop {} {\bibfield  {journal} {\bibinfo  {journal} {Chromatographia}\ }\textbf {\bibinfo {volume} {17}},\ \bibinfo {pages} {253} (\bibinfo {year} {1983})}\BibitemShut {NoStop}%
\bibitem [{\citenamefont {Moutzouris}\ \emph {et~al.}(2014)\citenamefont {Moutzouris}, \citenamefont {Papamichael}, \citenamefont {Betsis}, \citenamefont {Stavrakas}, \citenamefont {Hloupis},\ and\ \citenamefont {Triantis}}]{moutzouris2014}%
  \BibitemOpen
  \bibfield  {author} {\bibinfo {author} {\bibfnamefont {K.}~\bibnamefont {Moutzouris}}, \bibinfo {author} {\bibfnamefont {M.}~\bibnamefont {Papamichael}}, \bibinfo {author} {\bibfnamefont {S.~C.}\ \bibnamefont {Betsis}}, \bibinfo {author} {\bibfnamefont {I.}~\bibnamefont {Stavrakas}}, \bibinfo {author} {\bibfnamefont {G.}~\bibnamefont {Hloupis}},\ and\ \bibinfo {author} {\bibfnamefont {D.}~\bibnamefont {Triantis}},\ }\bibfield  {title} {\bibinfo {title} {Refractive, dispersive and thermo-optic properties of twelve organic solvents in the visible and near-infrared},\ }\href@noop {} {\bibfield  {journal} {\bibinfo  {journal} {Applied Physics B}\ }\textbf {\bibinfo {volume} {116}},\ \bibinfo {pages} {617} (\bibinfo {year} {2014})}\BibitemShut {NoStop}%
\bibitem [{\citenamefont {Bertie}\ and\ \citenamefont {Lan}(1997)}]{bertie1997liquid}%
  \BibitemOpen
  \bibfield  {author} {\bibinfo {author} {\bibfnamefont {J.~E.}\ \bibnamefont {Bertie}}\ and\ \bibinfo {author} {\bibfnamefont {Z.}~\bibnamefont {Lan}},\ }\bibfield  {title} {\bibinfo {title} {Liquid water- acetonitrile mixtures at 25 c: the hydrogen-bonded structure studied through infrared absolute integrated absorption intensities},\ }\href@noop {} {\bibfield  {journal} {\bibinfo  {journal} {The Journal of Physical Chemistry B}\ }\textbf {\bibinfo {volume} {101}},\ \bibinfo {pages} {4111} (\bibinfo {year} {1997})}\BibitemShut {NoStop}%
\end{thebibliography}%

\newpage
\begin{figure*}
\includegraphics[width=1\textwidth]{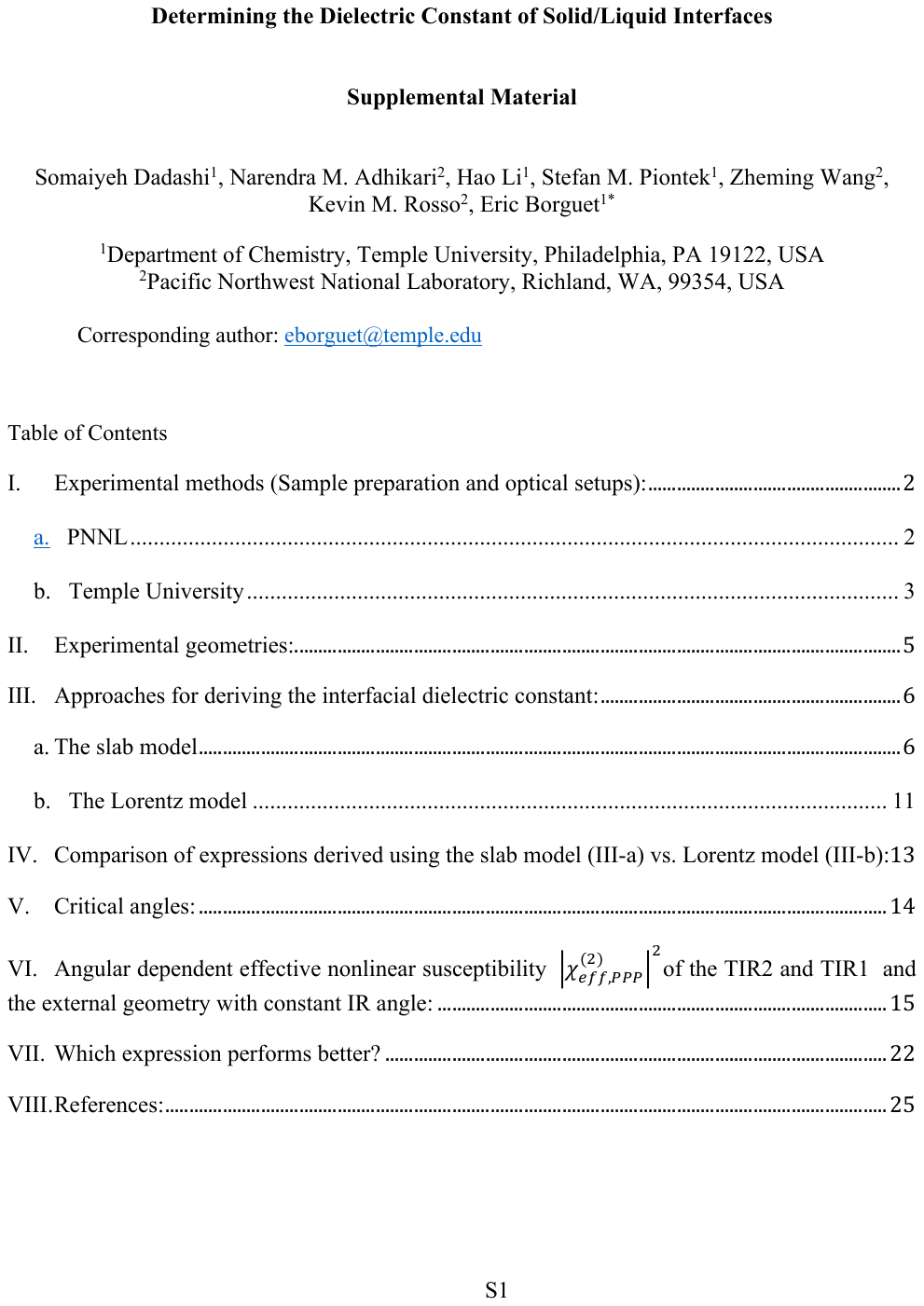}% Here is how to import EPS art
\end{figure*}
\begin{figure*}
\includegraphics[width=1\textwidth]{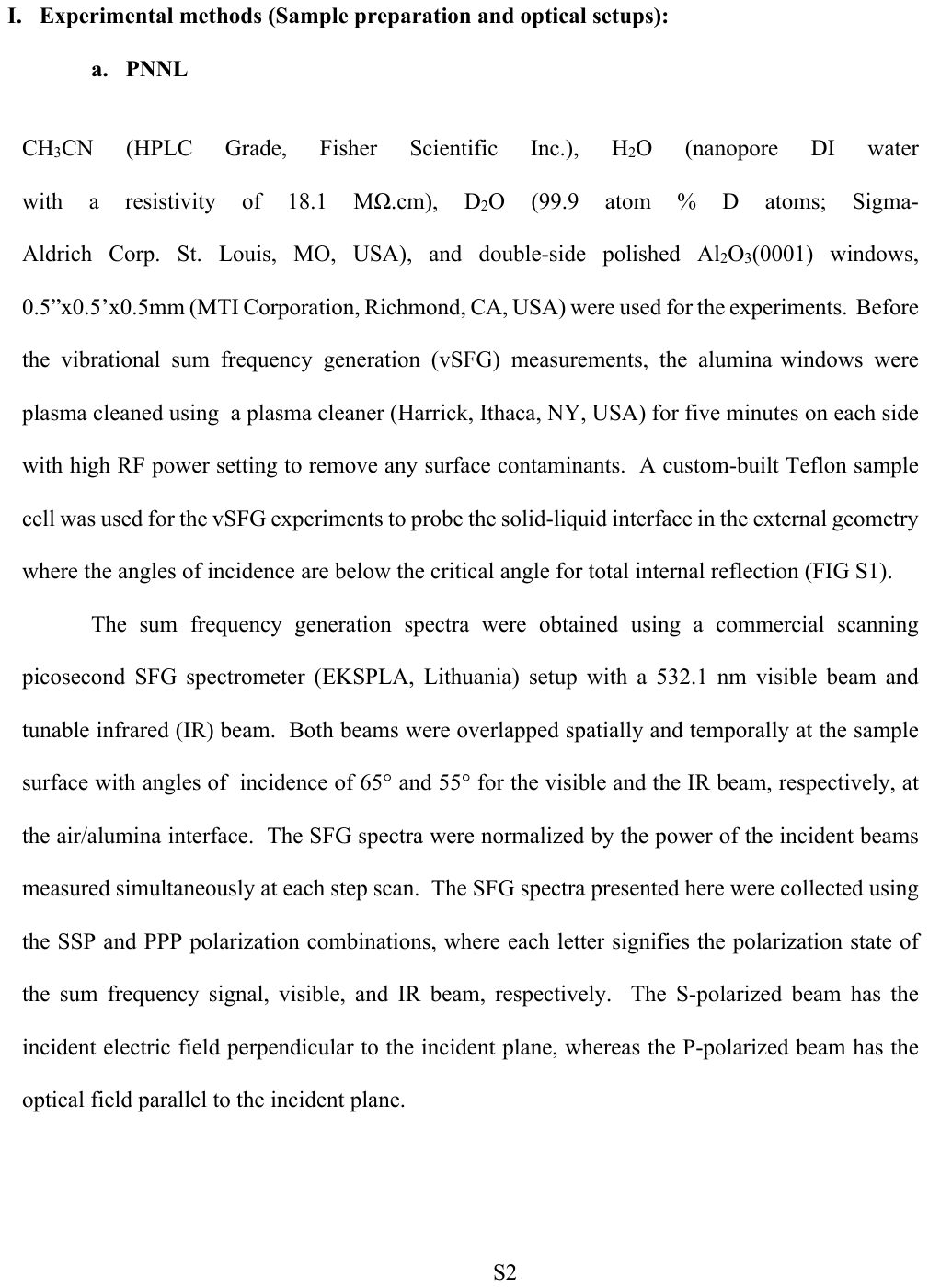}% Here is how to import EPS art
\end{figure*}
\begin{figure*}
\includegraphics[width=1\textwidth]{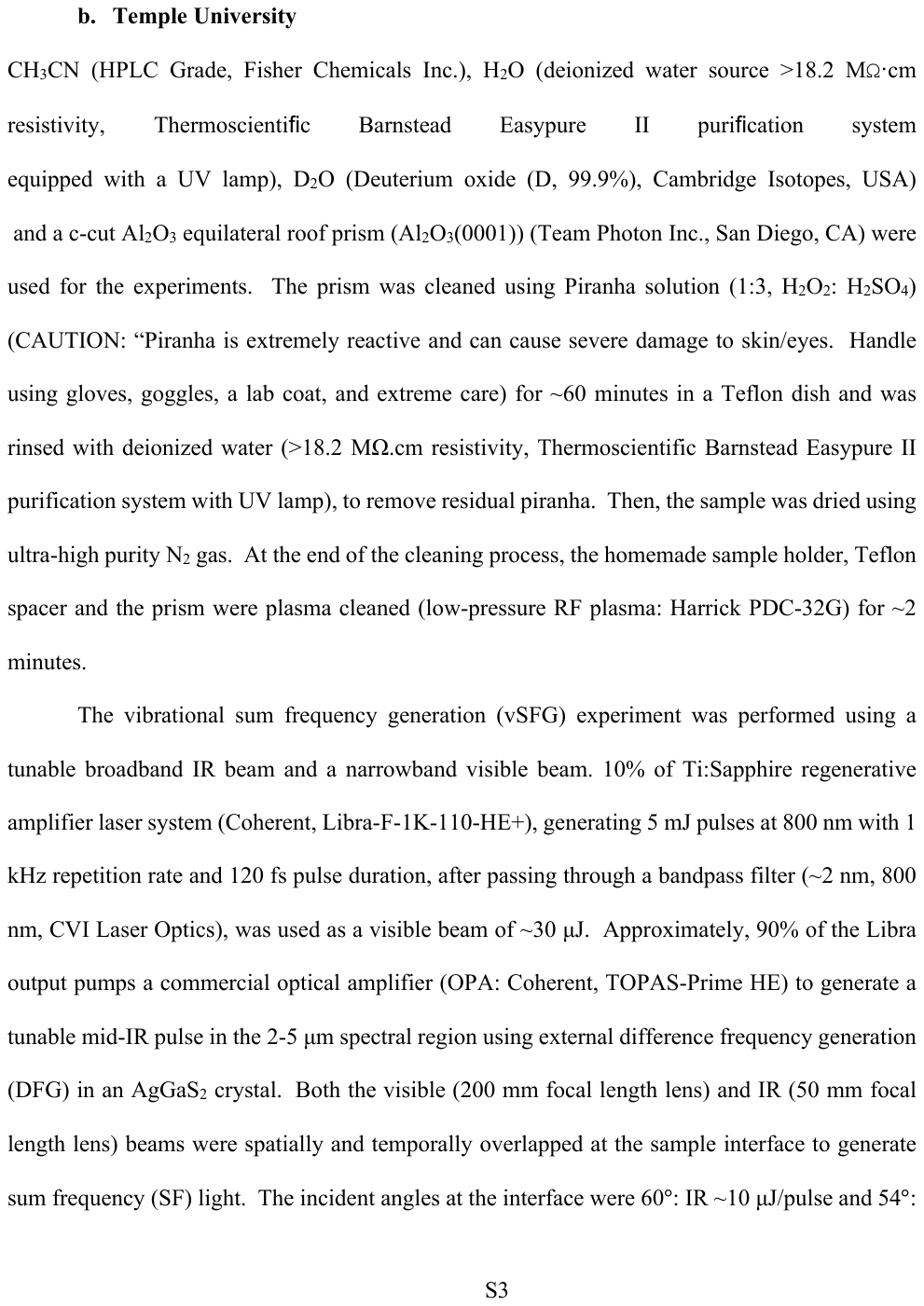}% Here is how to import EPS art
\end{figure*}
\begin{figure*}
\includegraphics[width=1\textwidth]{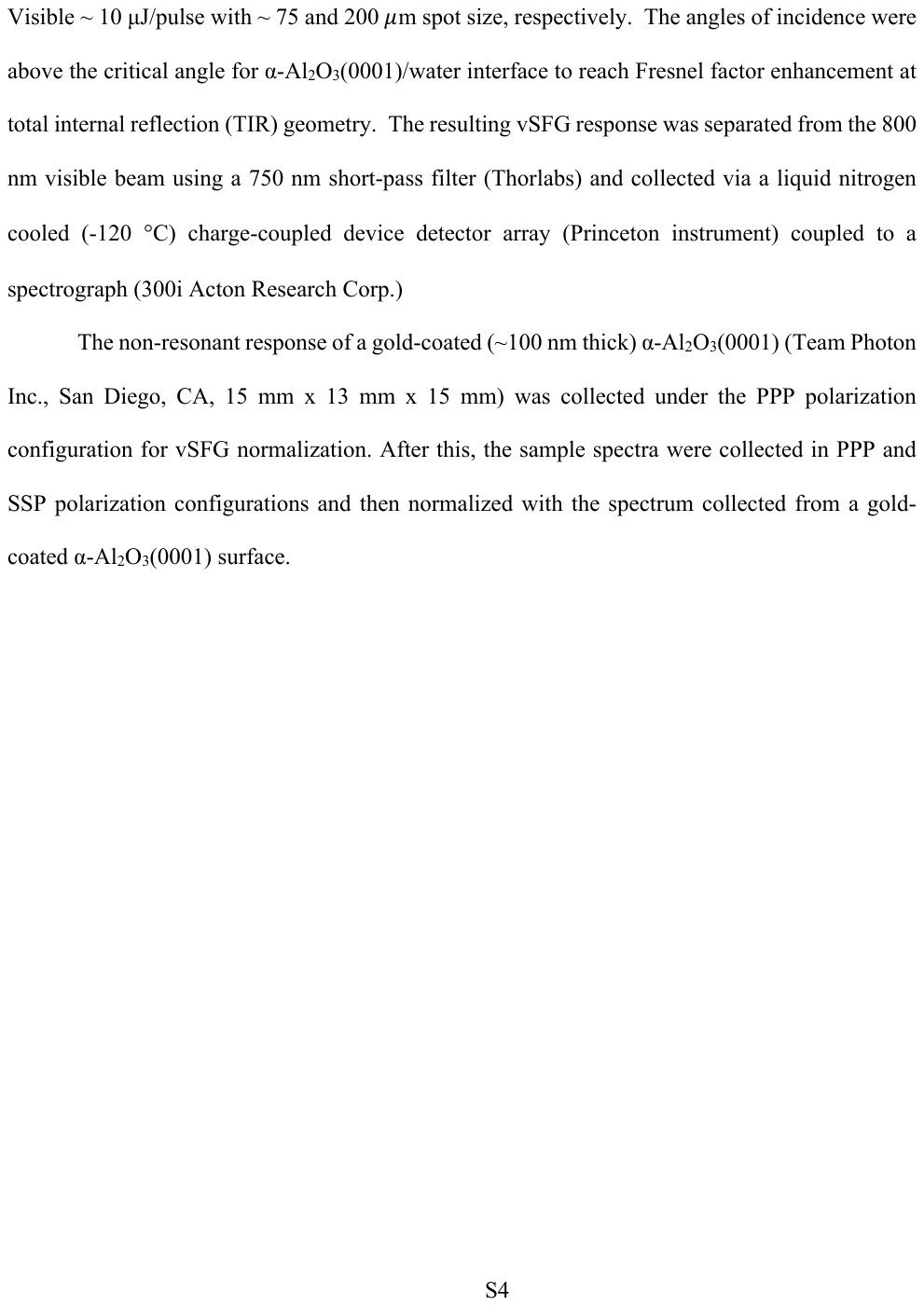}% Here is how to import EPS art
\end{figure*}
\begin{figure*}
\includegraphics[width=1\textwidth]{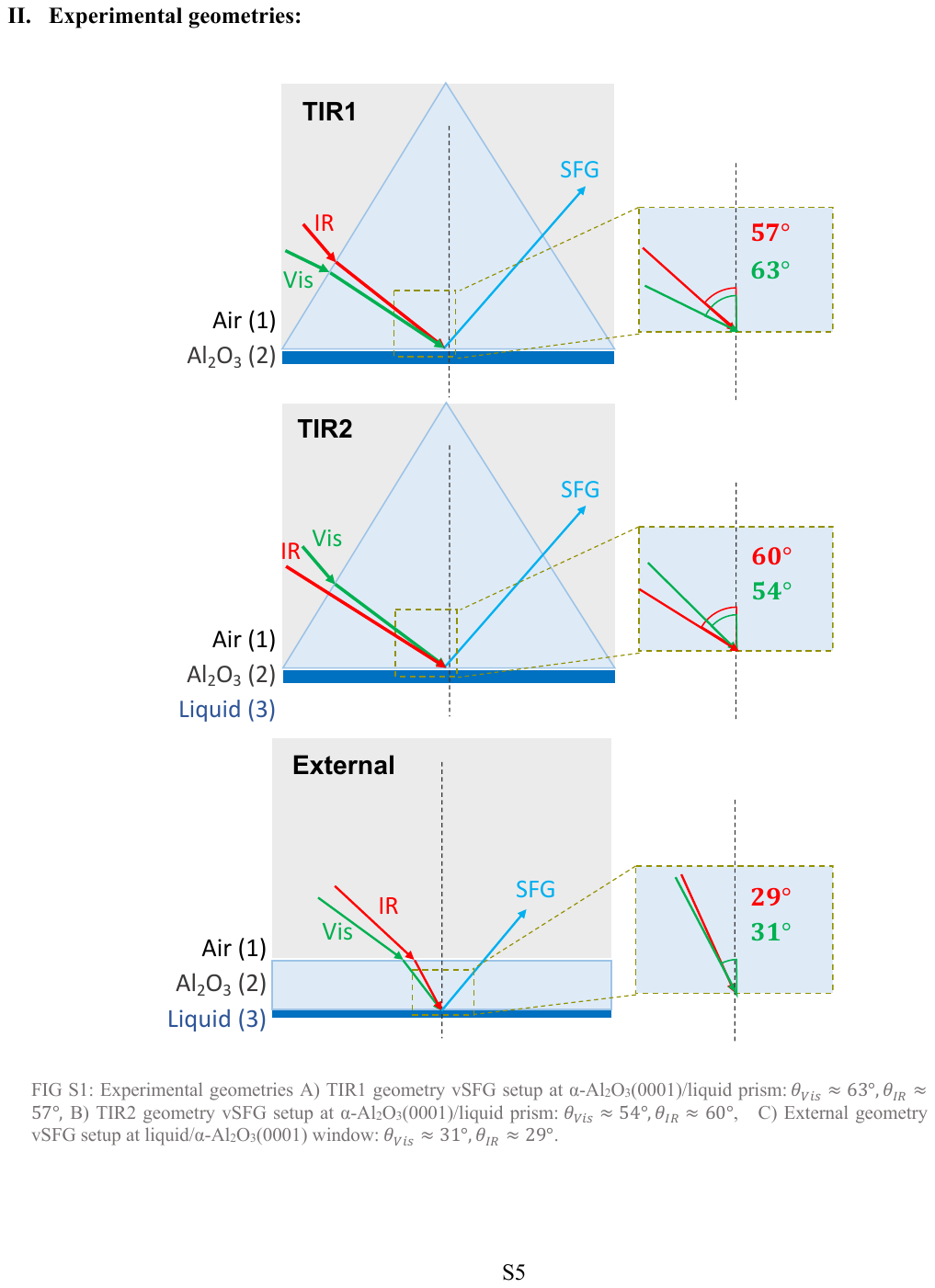}% Here is how to import EPS art
\end{figure*}
\begin{figure*}
\includegraphics[width=1\textwidth]{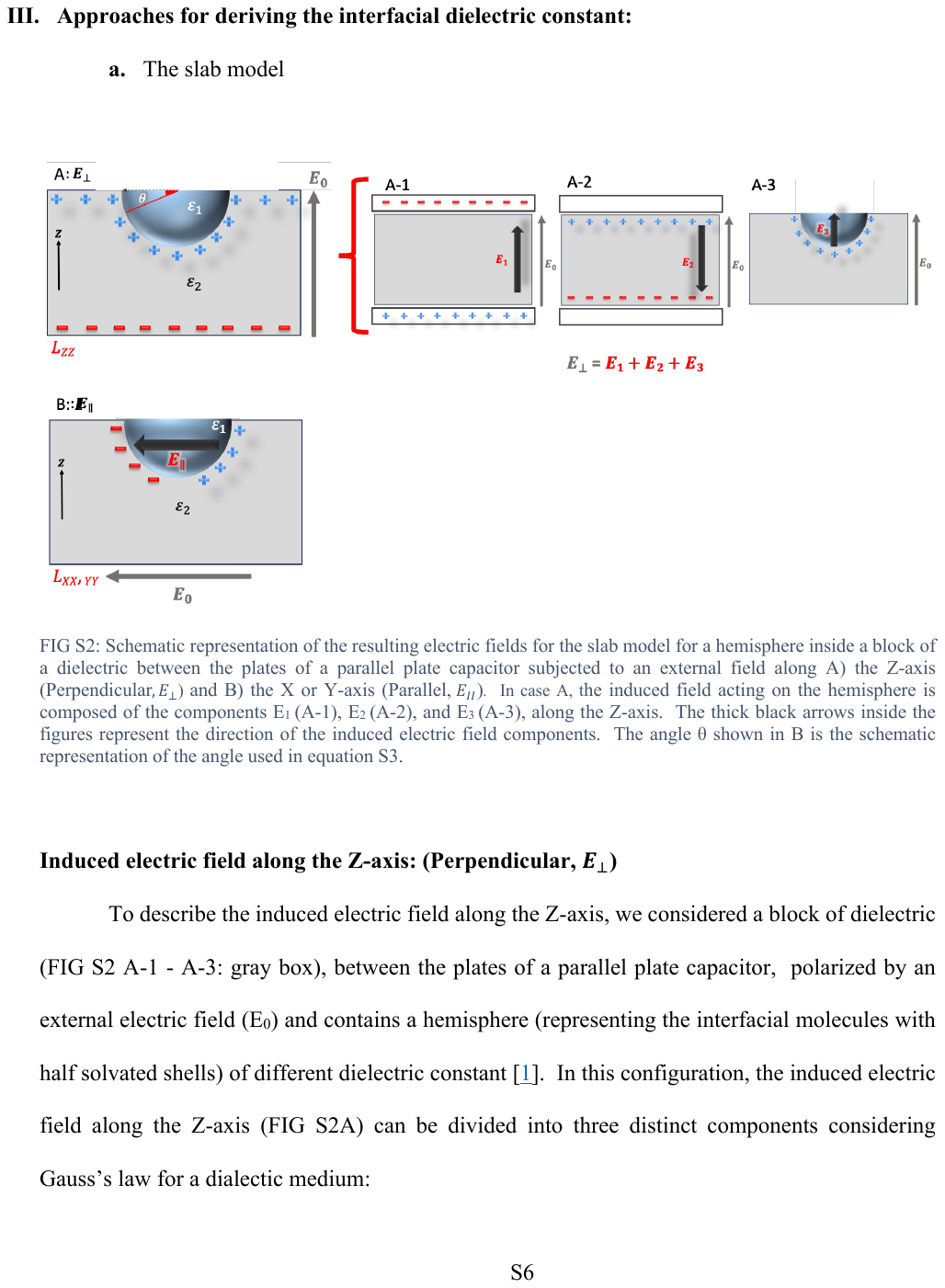}% Here is how to import EPS art
\end{figure*}
\begin{figure*}
\includegraphics[width=1\textwidth]{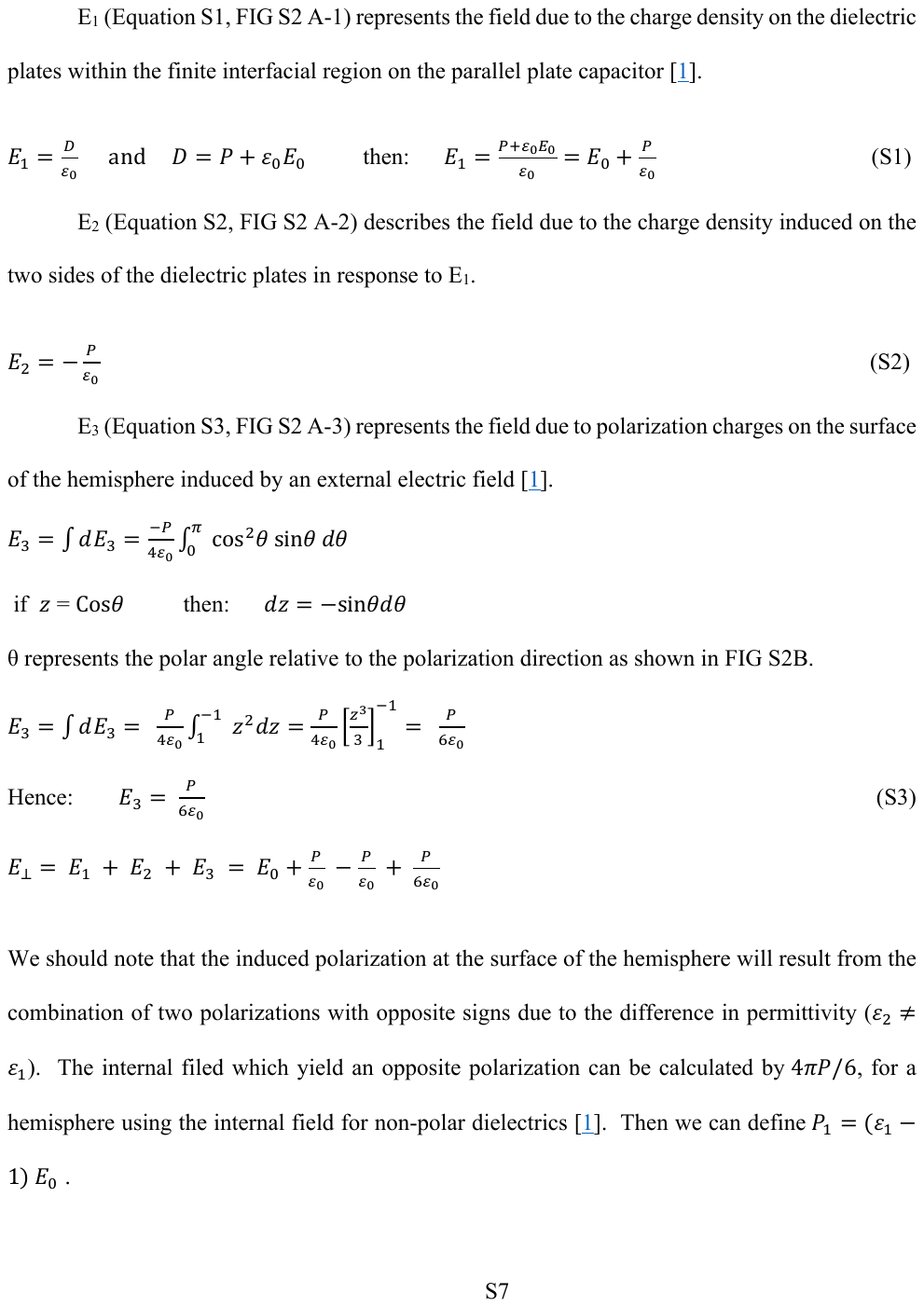}% Here is how to import EPS art
\end{figure*}
\begin{figure*}
\includegraphics[width=1\textwidth]{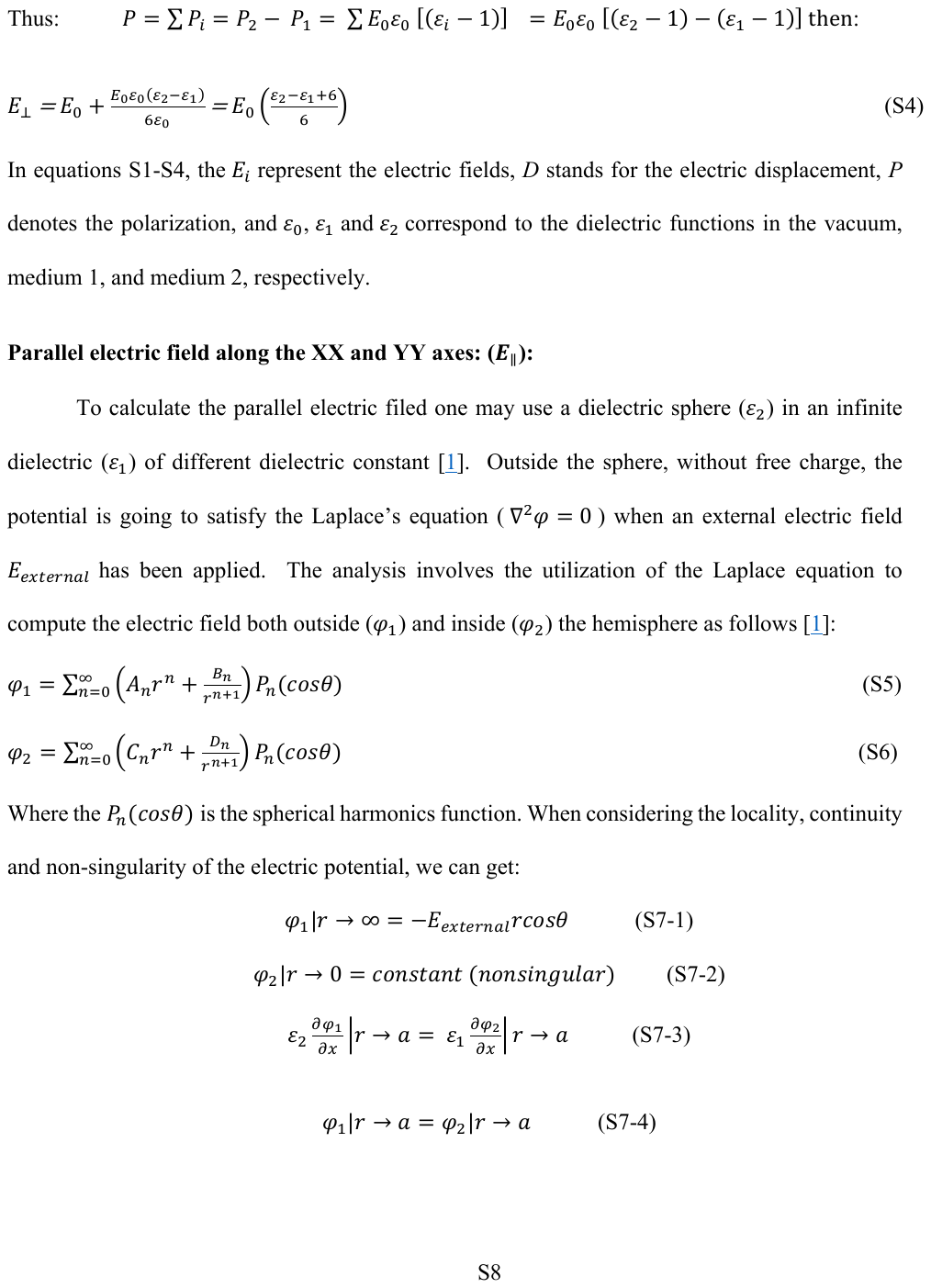}% Here is how to import EPS art
\end{figure*}
\begin{figure*}
\includegraphics[width=1\textwidth]{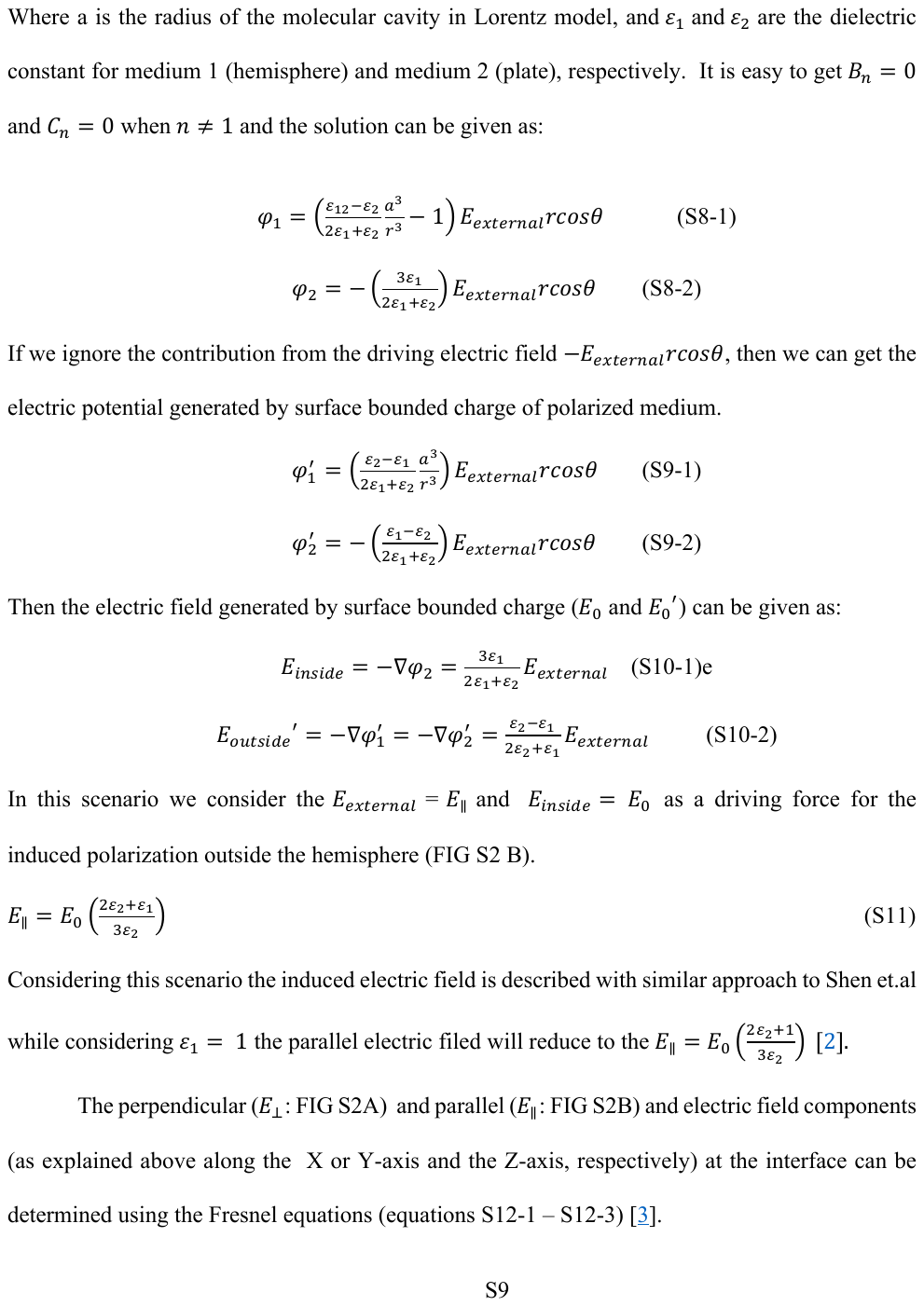}% Here is how to import EPS art
\end{figure*}
\begin{figure*}
\includegraphics[width=1\textwidth]{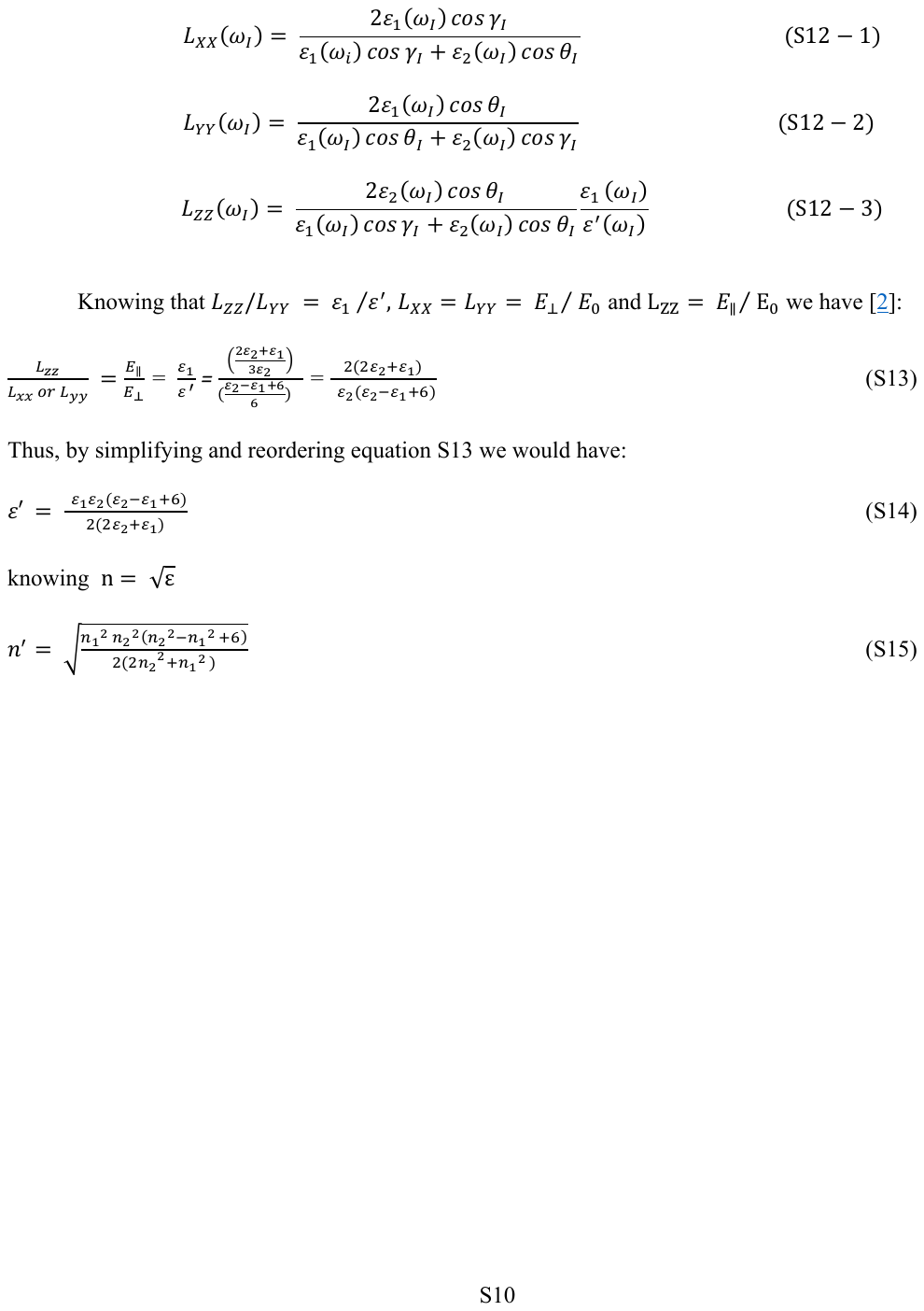}% Here is how to import EPS art
\end{figure*}
\begin{figure*}
\includegraphics[width=1\textwidth]{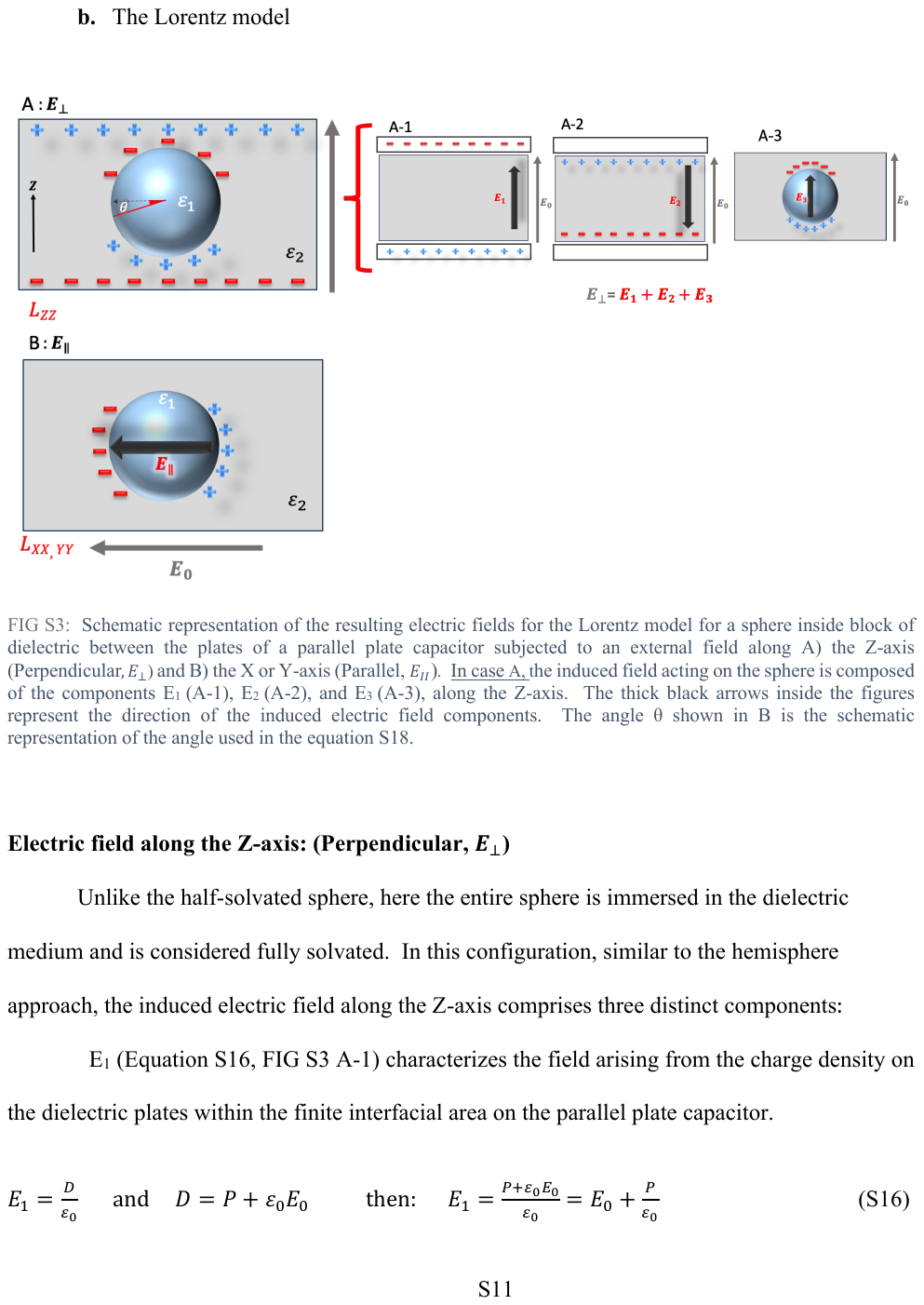}% Here is how to import EPS art
\end{figure*}
\begin{figure*}
\includegraphics[width=1\textwidth]{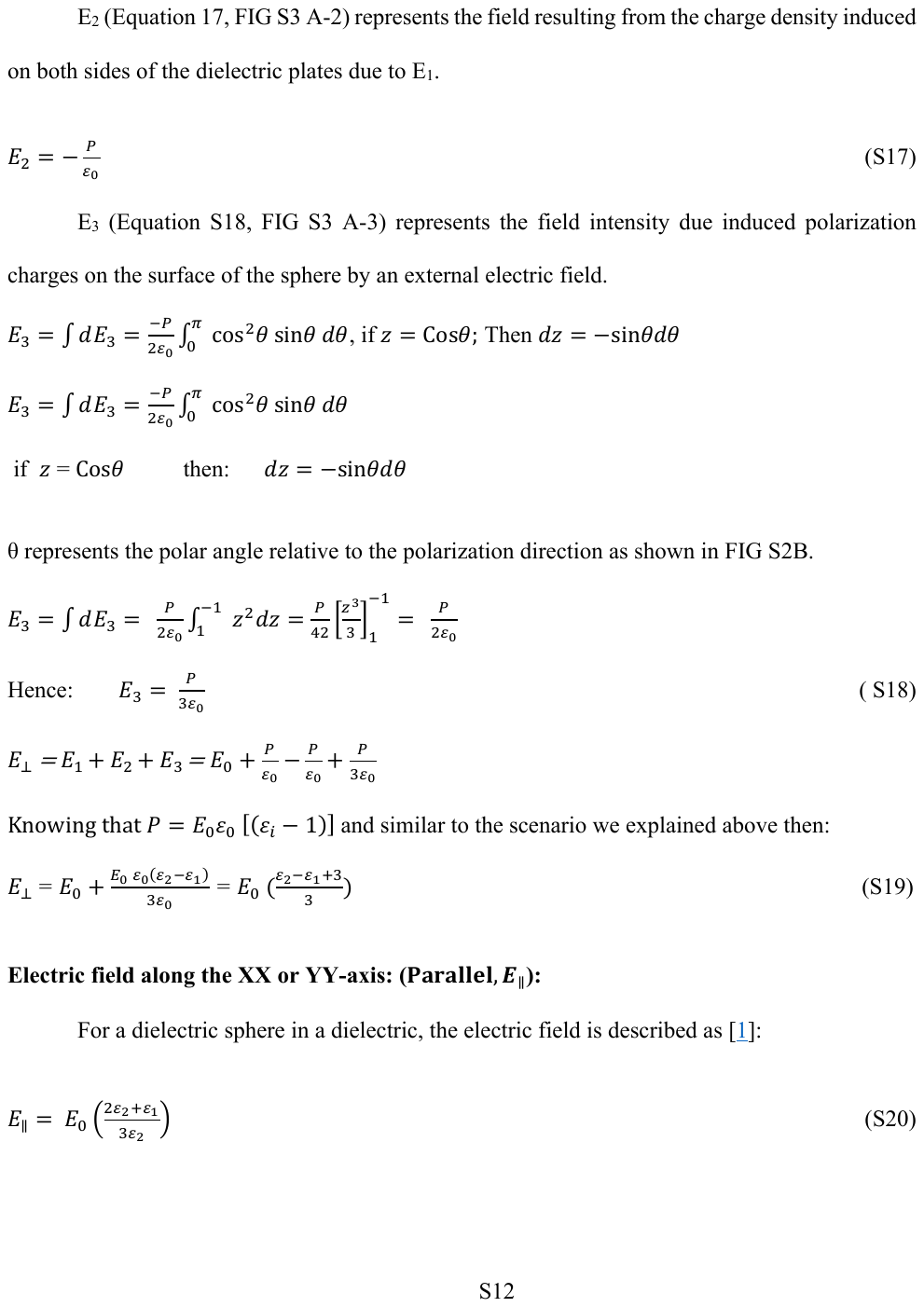}% Here is how to import EPS art
\end{figure*}
\begin{figure*}
\includegraphics[width=1\textwidth]{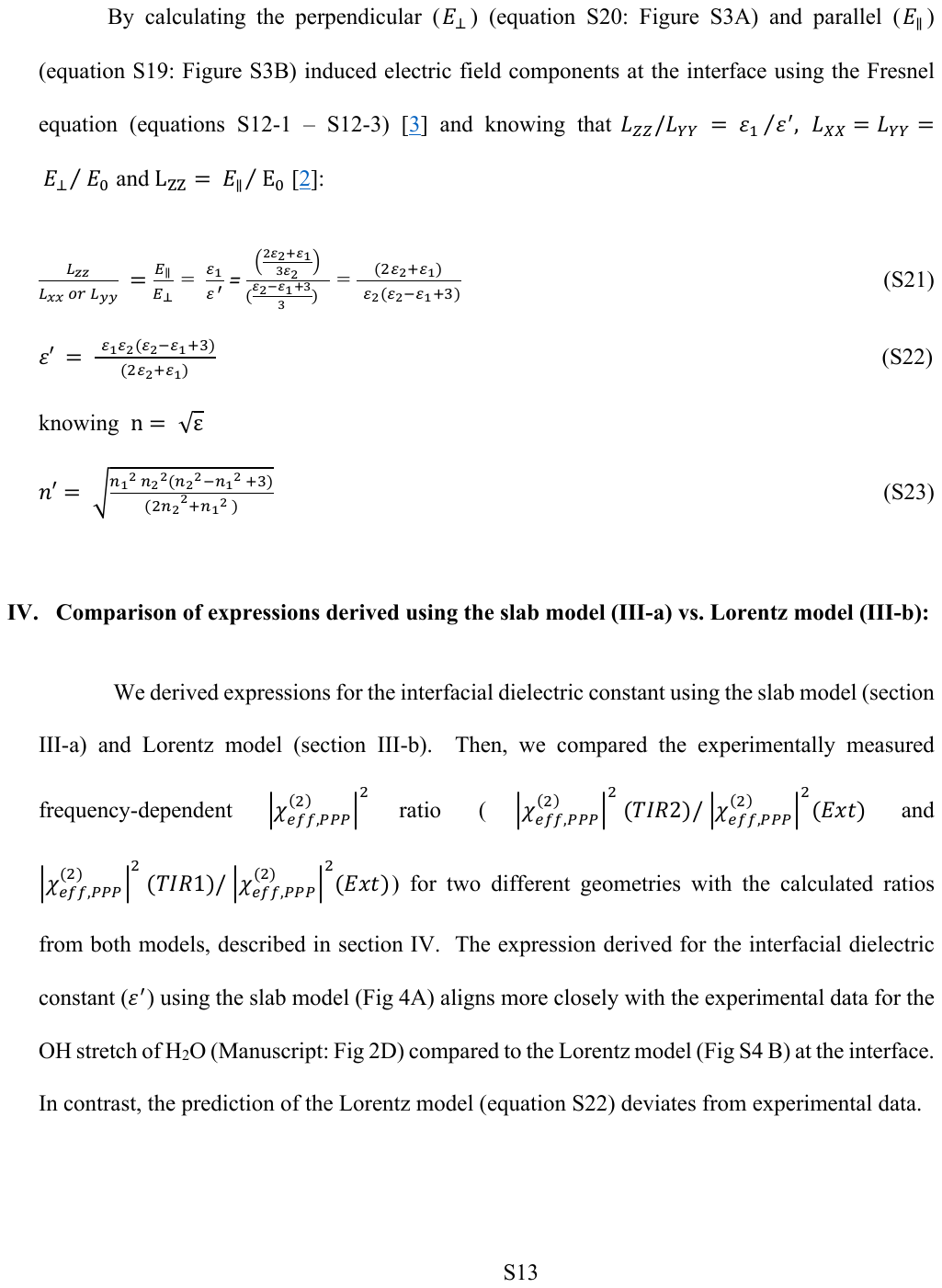}% Here is how to import EPS art
\end{figure*}
\begin{figure*}
\includegraphics[width=1\textwidth]{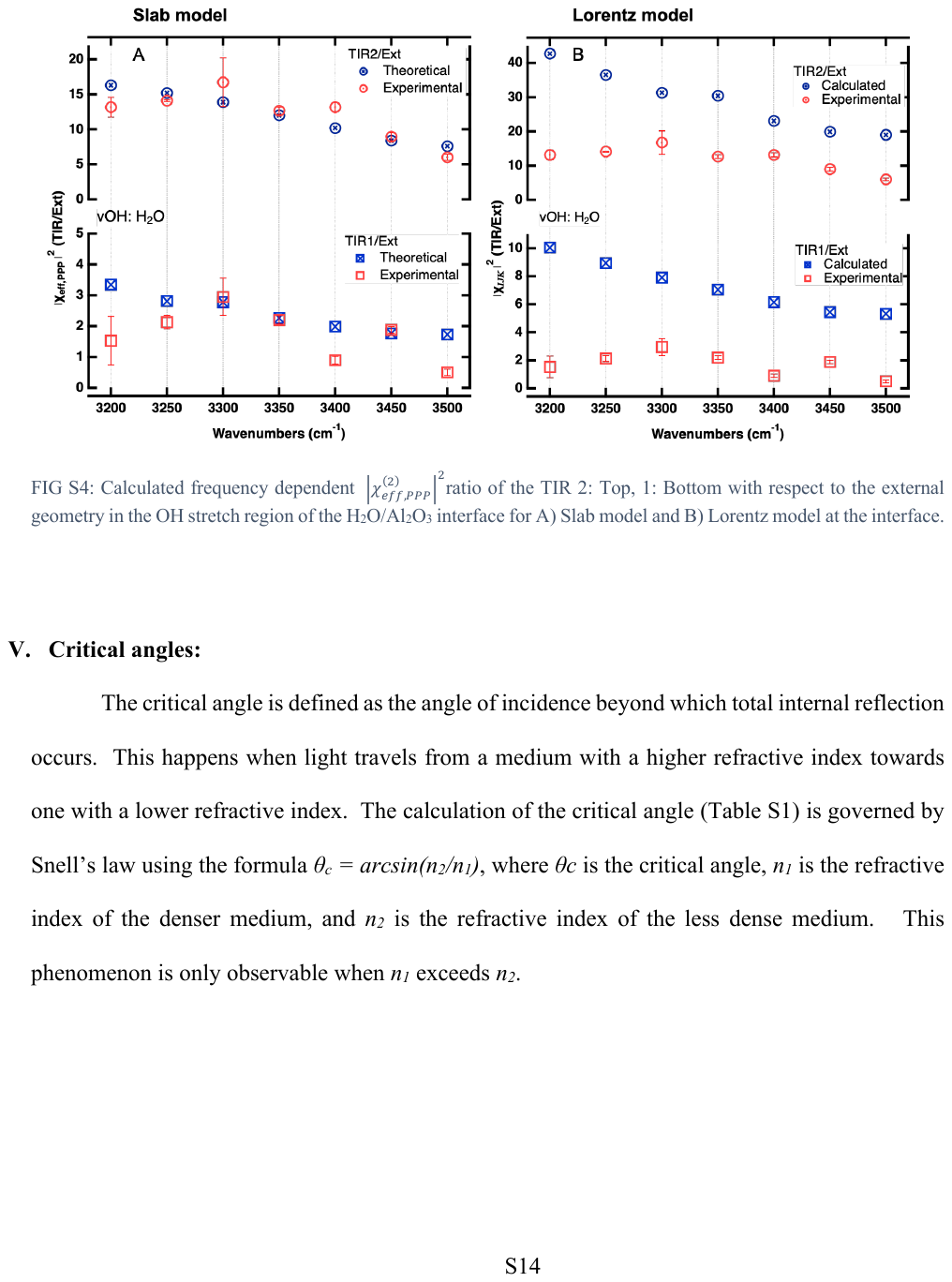}% Here is how to import EPS art
\end{figure*}
\begin{figure*}
\includegraphics[width=1\textwidth]{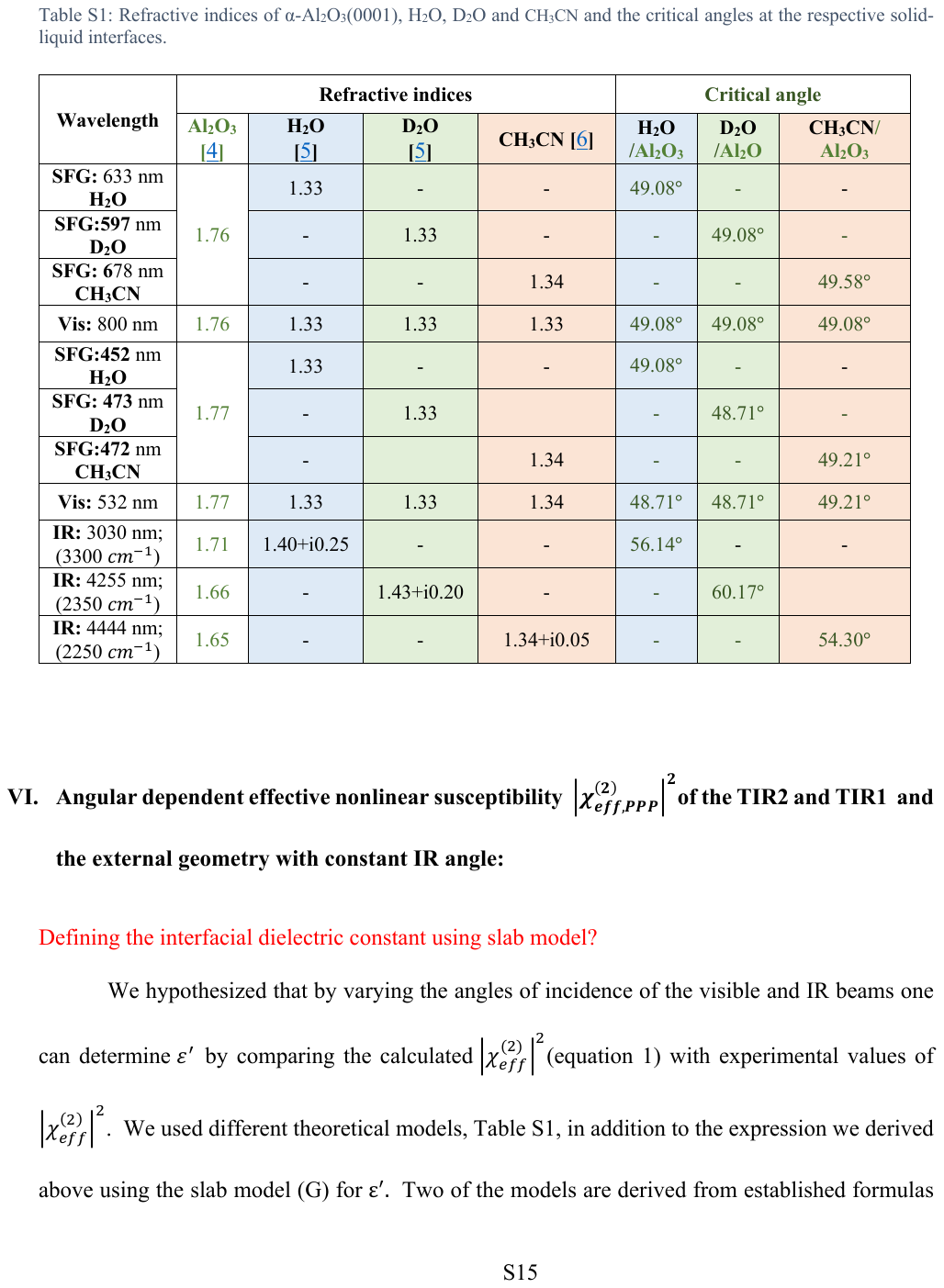}% Here is how to import EPS art
\end{figure*}
\begin{figure*}
\includegraphics[width=1\textwidth]{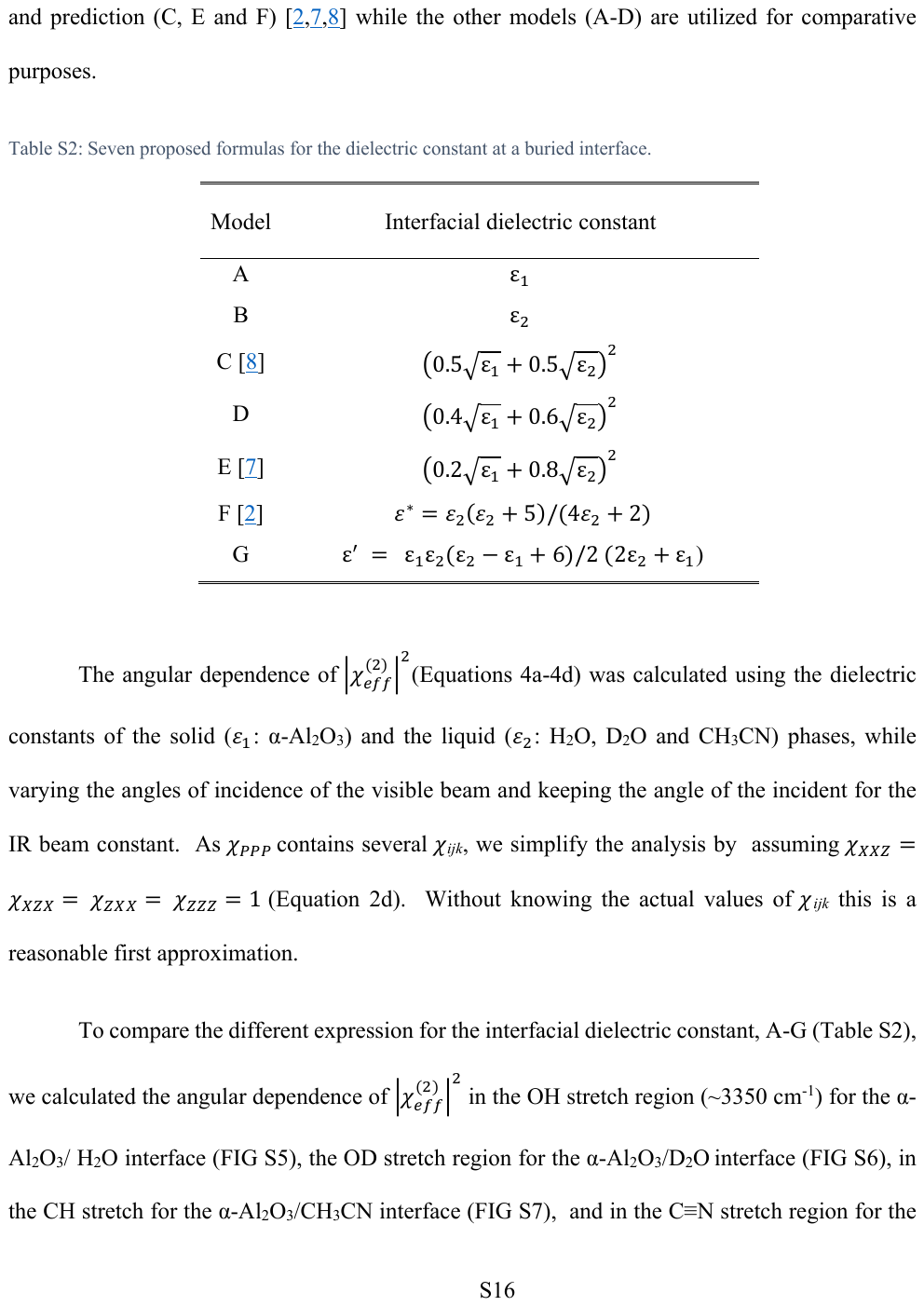}% Here is how to import EPS art
\end{figure*}
\begin{figure*}
\includegraphics[width=1\textwidth]{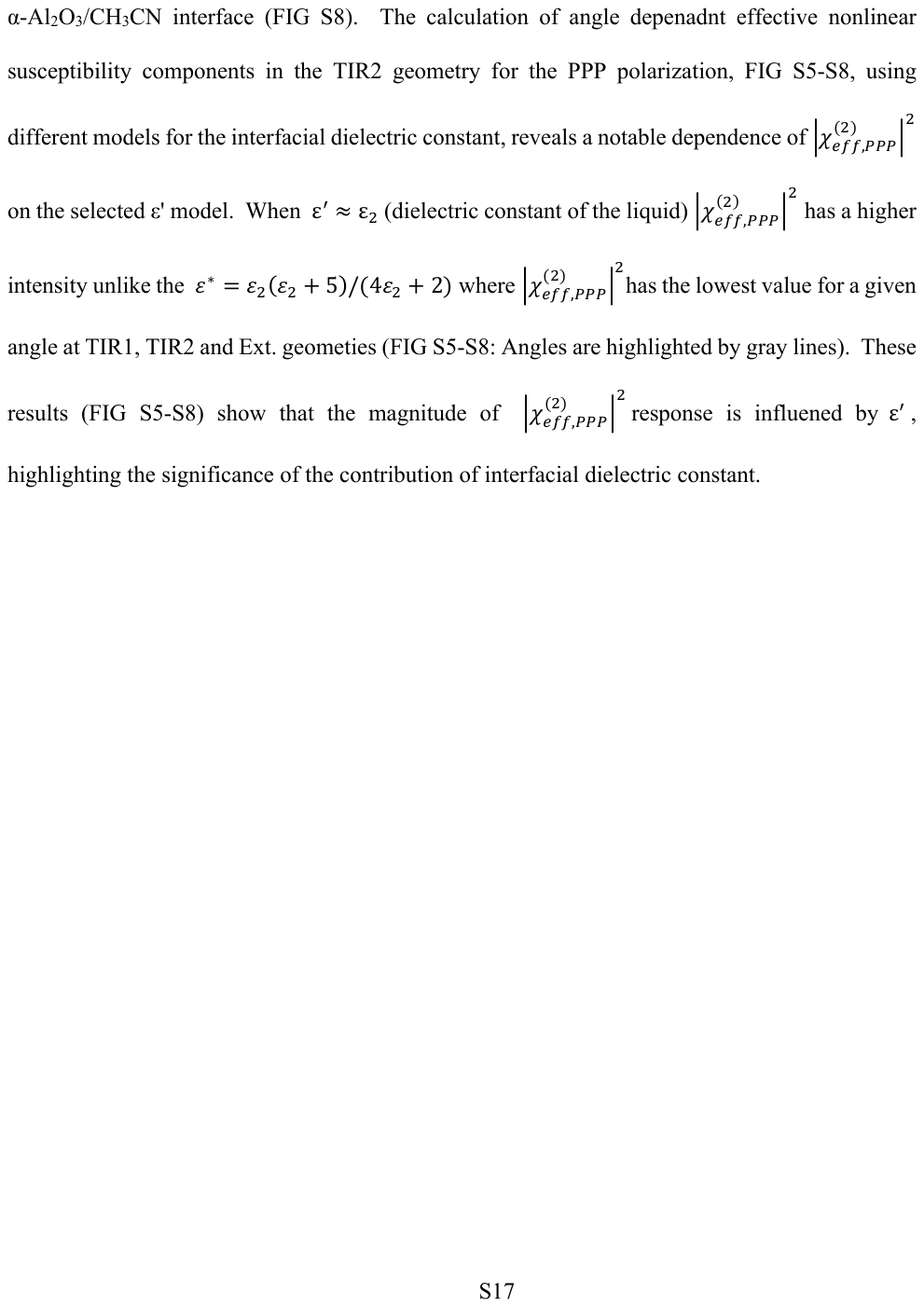}% Here is how to import EPS art
\end{figure*}
\begin{figure*}
\includegraphics[width=1\textwidth]{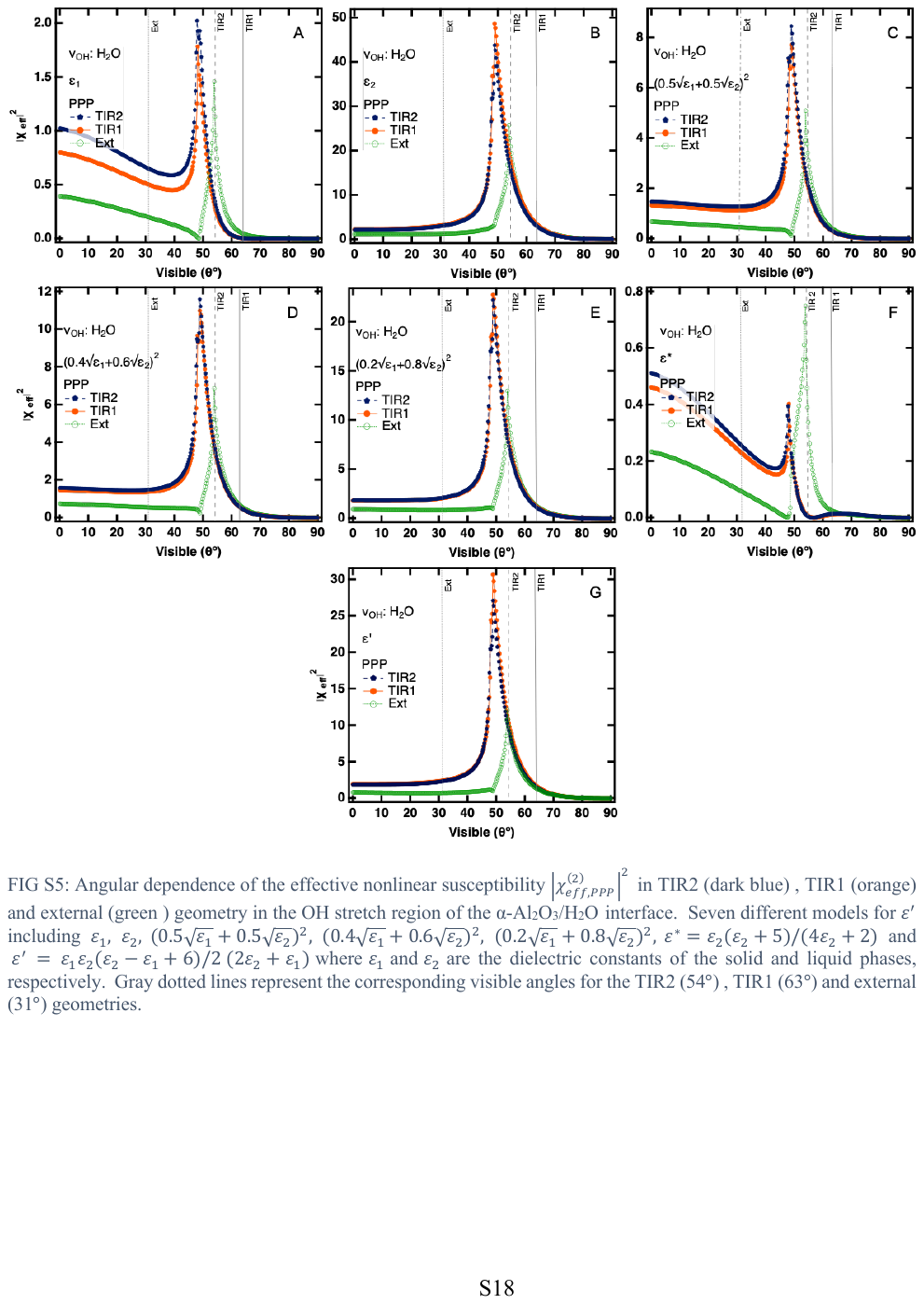}% Here is how to import EPS art
\end{figure*}
\begin{figure*}
\includegraphics[width=1\textwidth]{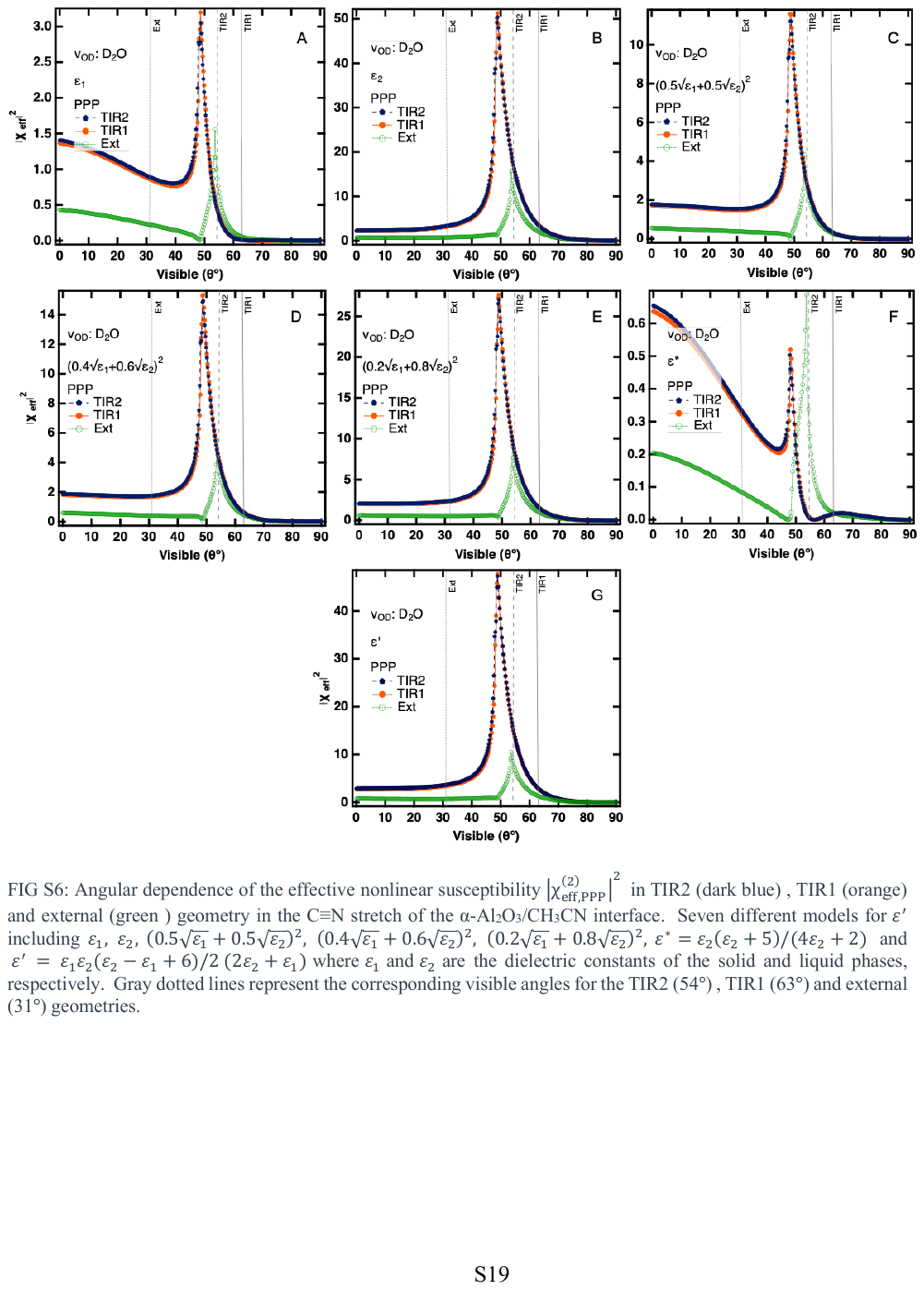}% Here is how to import EPS art
\end{figure*}
\begin{figure*}
\includegraphics[width=1\textwidth]{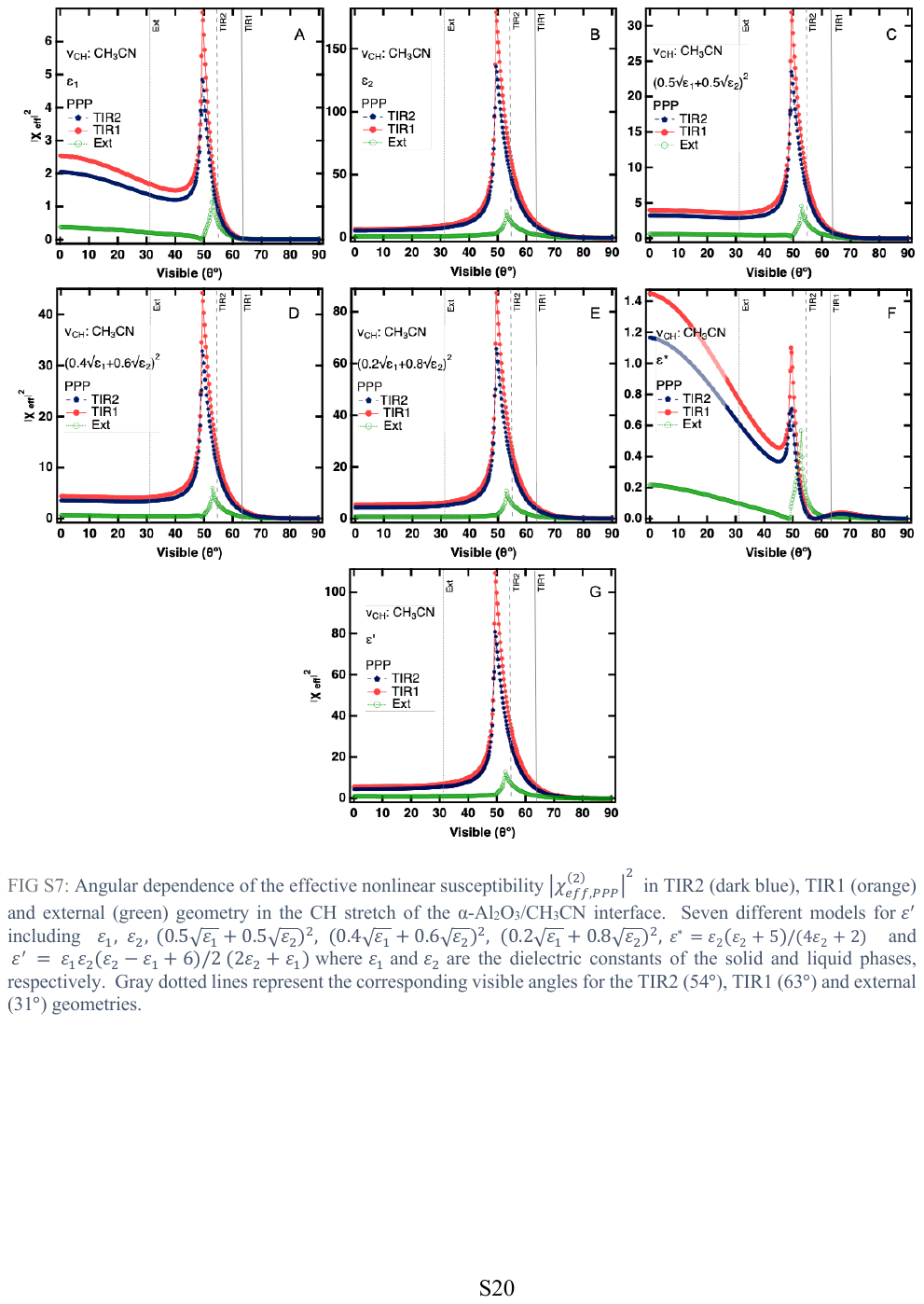}% Here is how to import EPS art
\end{figure*}
\begin{figure*}
\includegraphics[width=1\textwidth]{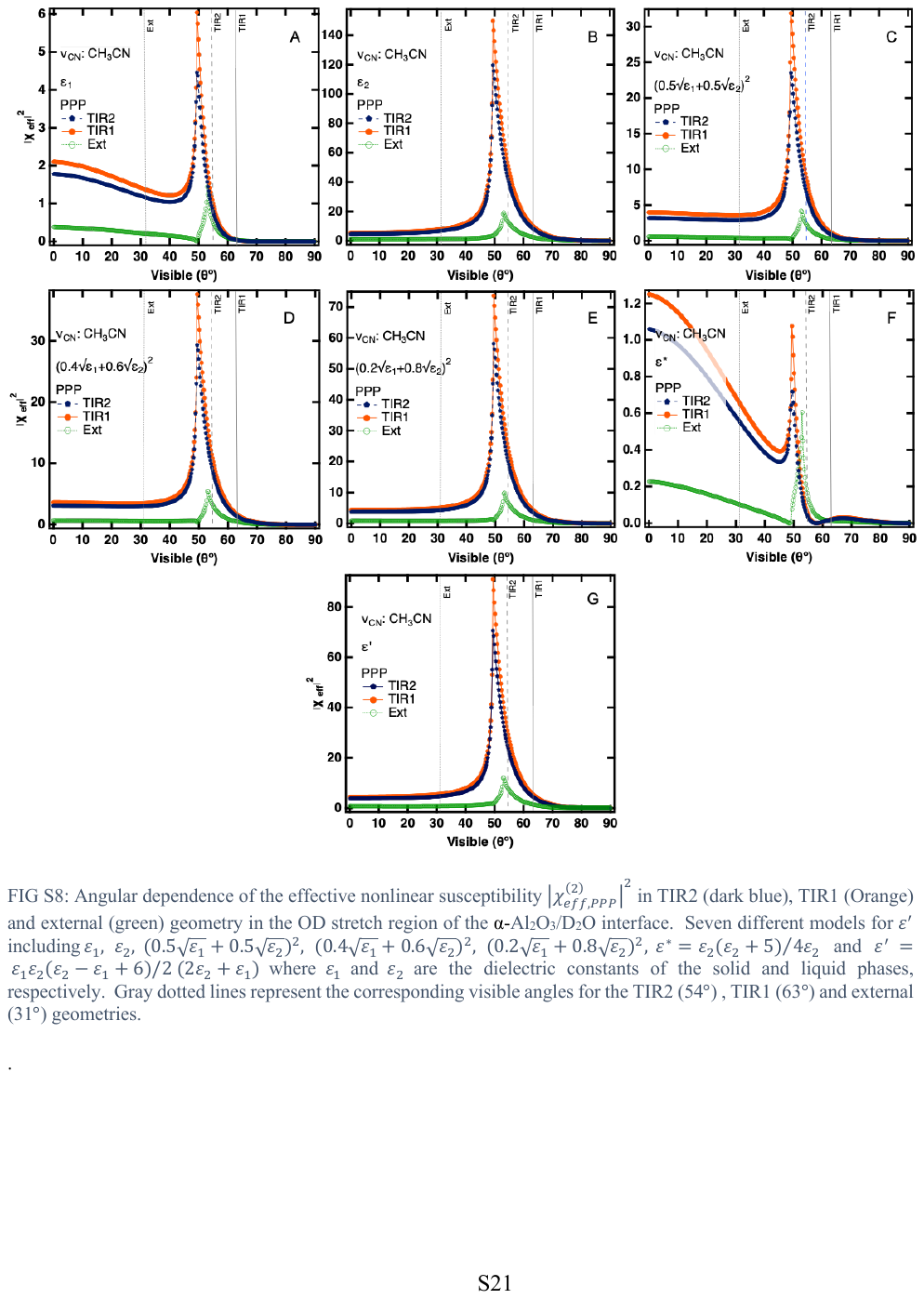}% Here is how to import EPS art
\end{figure*}
\begin{figure*}
\includegraphics[width=1\textwidth]{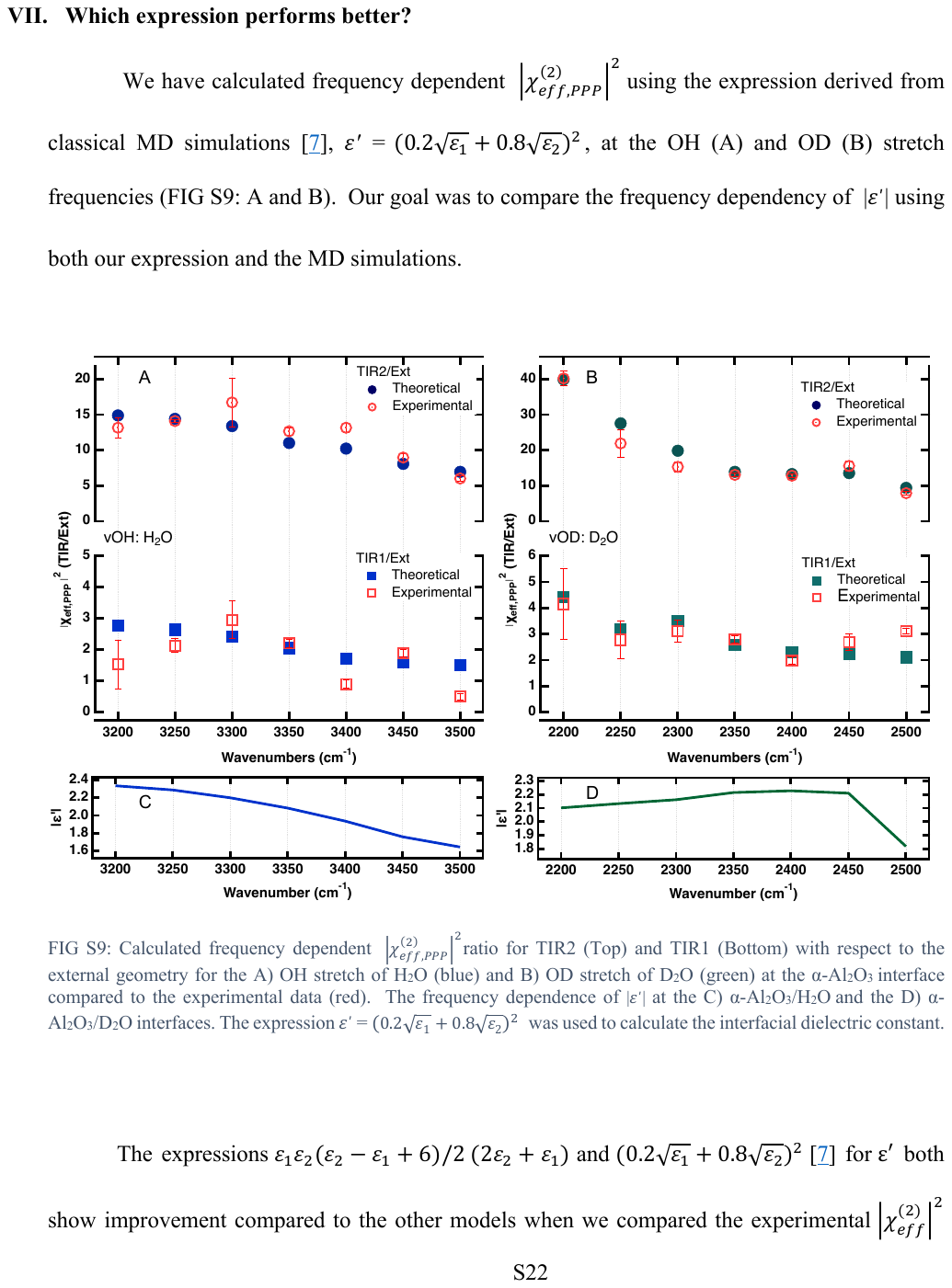}% Here is how to import EPS art
\end{figure*}
\begin{figure*}
\includegraphics[width=1\textwidth]{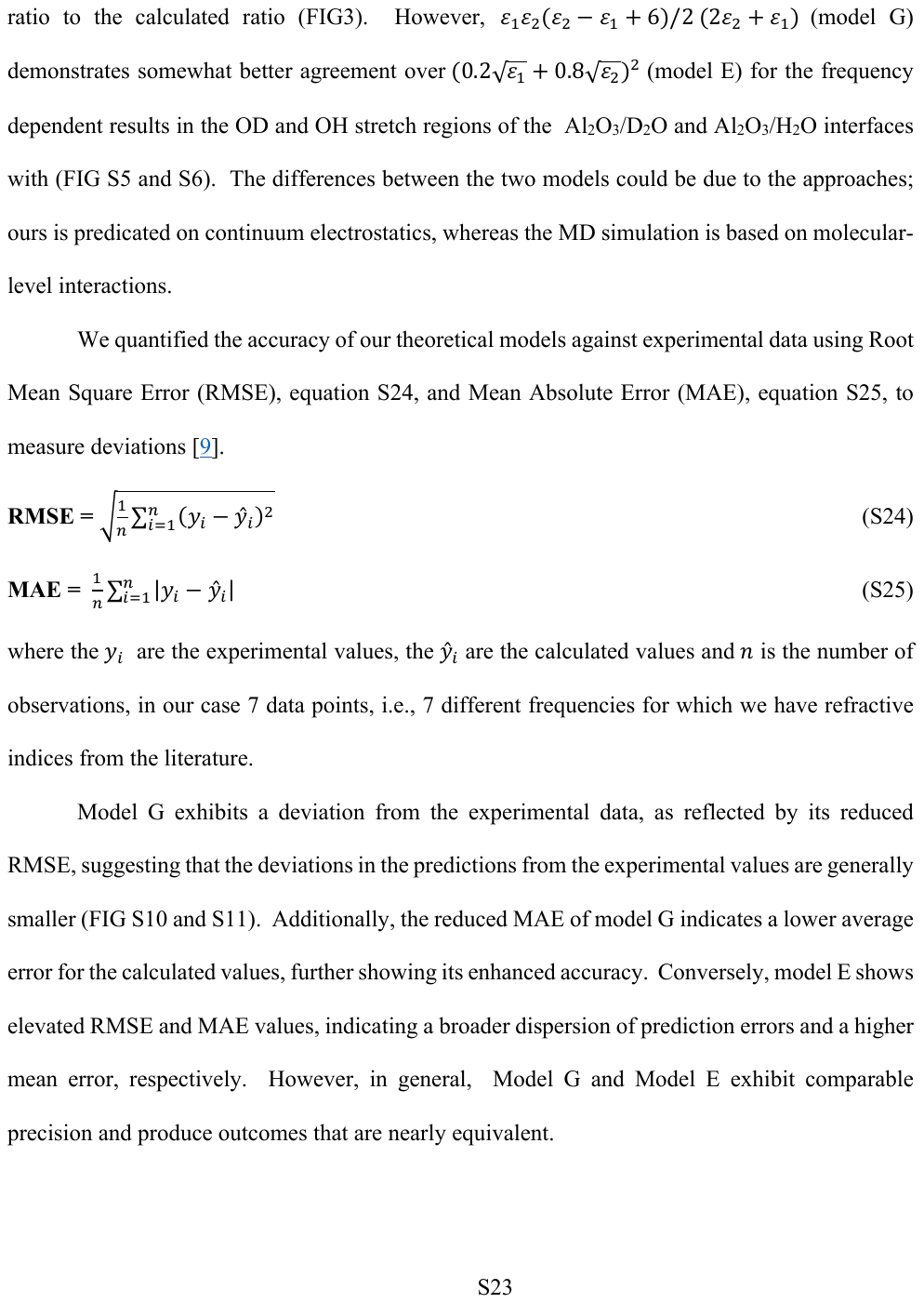}% Here is how to import EPS art
\end{figure*}
\begin{figure*}
\includegraphics[width=1\textwidth]{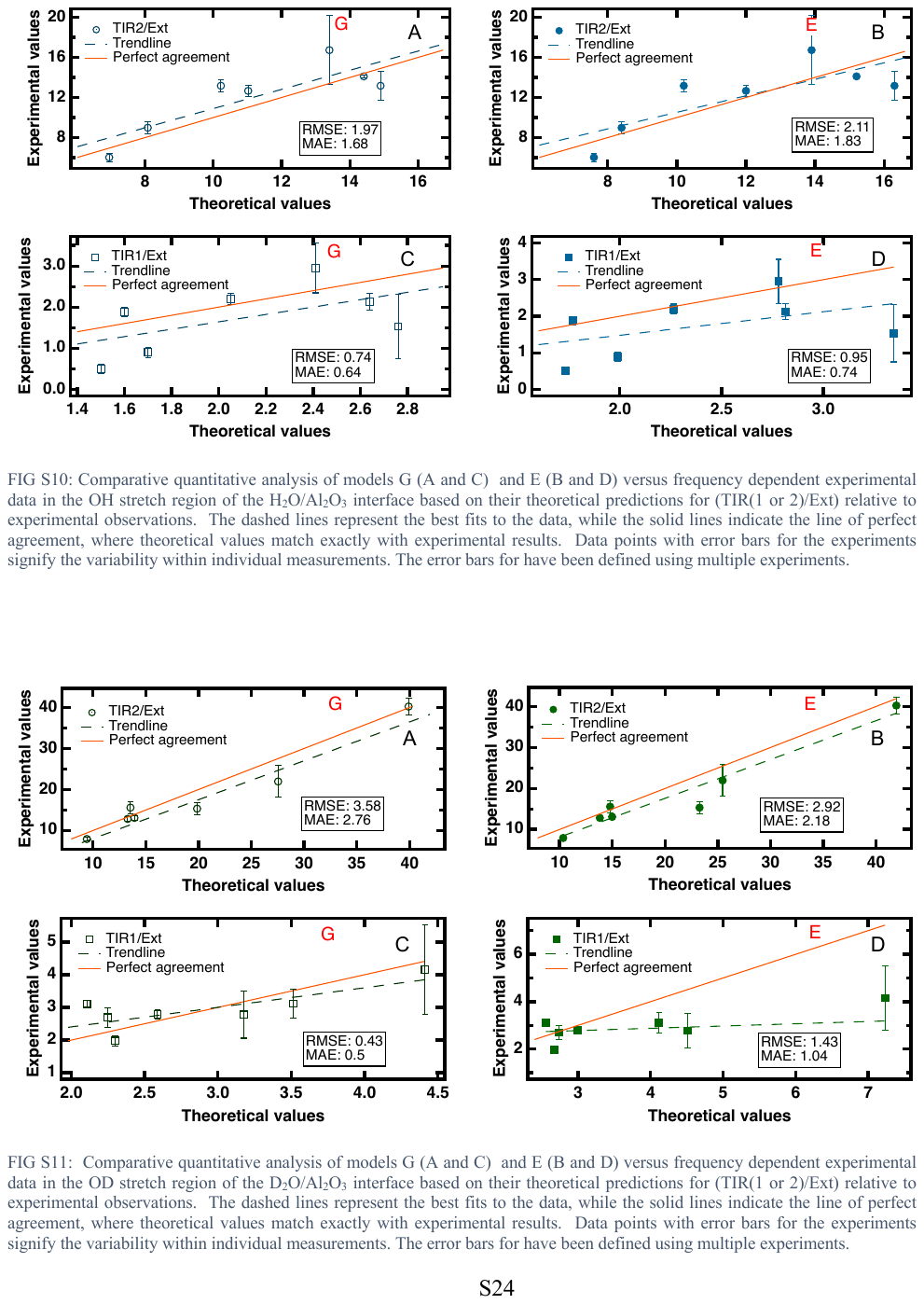}% Here is how to import EPS art
\end{figure*}
\begin{figure*}
\includegraphics[width=1\textwidth]{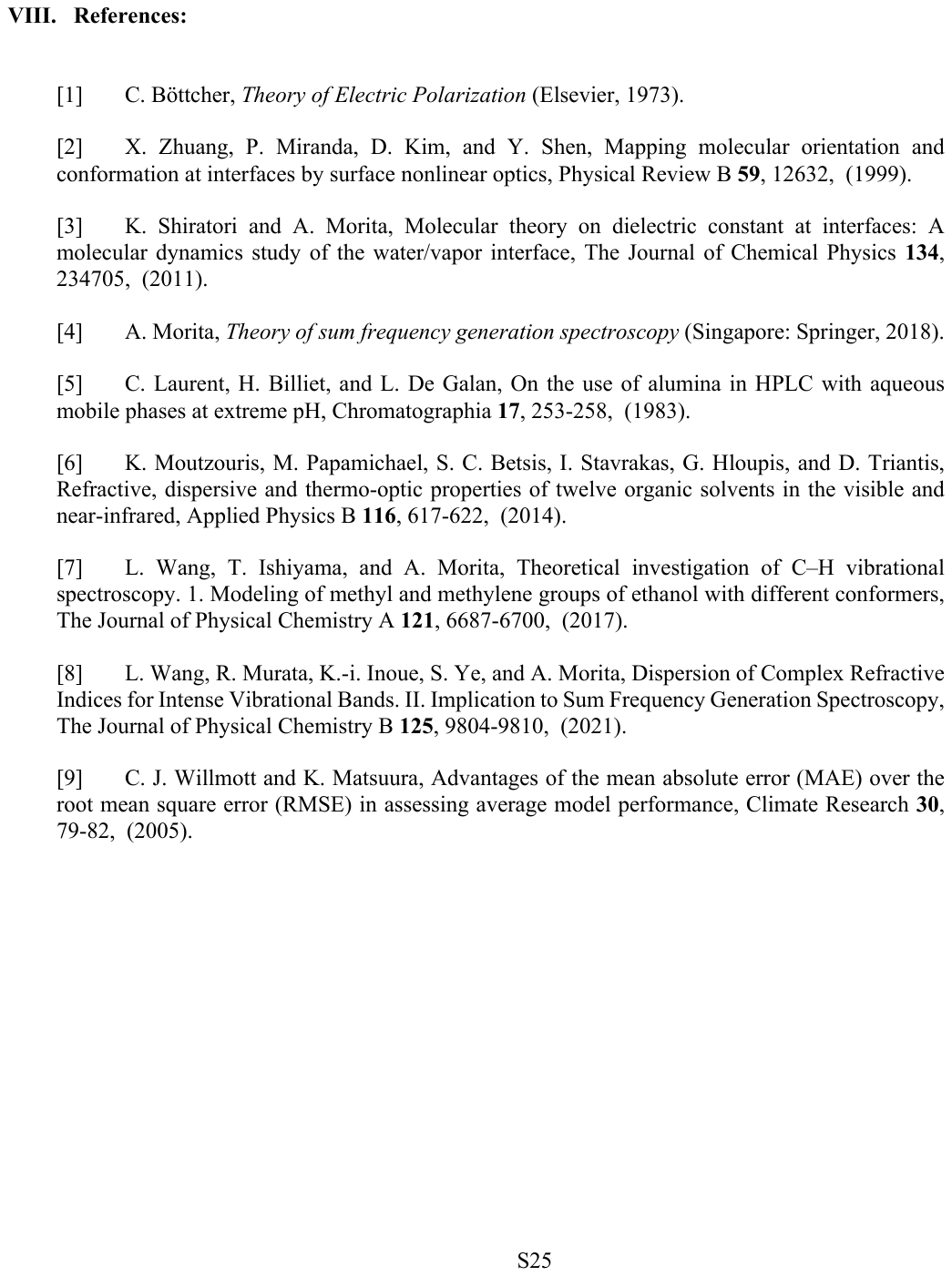}% Here is how to import EPS art
\end{figure*}

\end{document}